\documentclass[prb,a4paper,showpacs,twocolumn,superscriptaddress,longbibliography]{revtex4-1}

\usepackage{amssymb}
\usepackage{amsmath}
\usepackage{amsfonts}
\usepackage{graphicx}
\usepackage{bm}
\usepackage{color}
\usepackage{multirow}
\usepackage{natbib}
\usepackage{hyperref}

\DeclareMathOperator{\Tr}{Tr}
\DeclareMathOperator{\spp}{sp}
\DeclareMathOperator{\sgn}{sgn}
\DeclareMathOperator{\liq}{li_2}

\DeclareMathOperator{\im}{Im}
\DeclareMathOperator{\re}{Re}

\begin{document}

\title{Local density of states and its mesoscopic fluctuations  near the transition to a superconducting state in disordered systems}

\author{I.S.~Burmistrov}

\affiliation{L.D.~Landau Institute for Theoretical Physics, Kosygina
  street 2, 117940 Moscow, Russia}

\affiliation{Moscow
Institute of Physics and Technology, 141700 Moscow, Russia}

\author{I.V.~Gornyi}
\affiliation{
 Institut f\"ur Nanotechnologie, Karlsruhe Institute of Technology,
 76021 Karlsruhe, Germany
}
\affiliation{
 A.F.~Ioffe Physico-Technical Institute,
 194021 St.~Petersburg, Russia.
}
\affiliation{
 \mbox{Institut f\"ur Theorie der kondensierten Materie,
 Karlsruhe Institute of Technology, 76128 Karlsruhe, Germany}
}
\affiliation{L.D.~Landau Institute for Theoretical Physics, Kosygina
  street 2, 117940 Moscow, Russia}

\author{A.D.~Mirlin}
\affiliation{
 Institut f\"ur Nanotechnologie, Karlsruhe Institute of Technology,
 76021 Karlsruhe, Germany
}
\affiliation{
 \mbox{Institut f\"ur Theorie der kondensierten Materie,
 Karlsruhe Institute of Technology, 76128 Karlsruhe, Germany}
}
\affiliation{
 Petersburg Nuclear Physics Institute,
 188300 St.~Petersburg, Russia.
}
\affiliation{L.D.~Landau Institute for Theoretical Physics, Kosygina
  street 2, 117940 Moscow, Russia}

\begin{abstract}
We develop a theory of the local density of states (LDOS) of disordered superconductors, employing the non-linear sigma-model formalism
and the renormalization-group framework. The theory takes into account the interplay of disorder and interaction couplings in
all channels, treating the systems with short-range and Coulomb interactions on equal footing.
We explore 2D systems that would be Anderson insulators in the absence of interaction and 2D or 3D systems that undergo
Anderson transition in the absence of interaction. We evaluate both the average tunneling density of states and its
mesoscopic fluctuations which are related to the LDOS multifractality in normal disordered systems.
The obtained average LDOS shows a pronounced depletion around the Fermi energy, both in the metallic phase (i.e., above the superconducting critical temperature $T_c$) and in the insulating phase near the superconductor-insulator transition (SIT).
The fluctuations of the LDOS are found to be particularly strong for the case of short-range interactions 
-- especially, in the regime when $T_c$  is enhanced by Anderson localization. On the other hand, the long-range
Coulomb repulsion reduces the mesoscopic LDOS fluctuations. However, also in a model with Coulomb interaction, the fluctuations become strong when the systems approaches the SIT.
\end{abstract}


\maketitle

\section{Introduction}
\label{s1}

Disordered superconductors show remarkable physics governed by interplay of superconductivity and Anderson localization. In particular, in two-dimensional (2D) systems, the competition between these two phenomena leads to a direct quantum phase transition between the insulating and superconducting states---the superconductor-insulator transition (SIT) [\onlinecite{Goldman98},\onlinecite{GantmakherDolgopolov2010}].
This is a zero-temperature transition that may be driven by varying the normal-state resistivity of a disordered film; experimentally, this is usually achieved by changing the film thickness.
At a finite (but sufficiently low) temperature the insulating and superconducting phases of the film are separated by a metallic state.
The physics of SIT and, more generally, of insulating, superconducting, and metallic states around SIT, has attracted a great deal of attention.

On the experimental side, two complementary approaches have been widely used to characterize the physics of disordered superconducting films under variation of temperature, film thickness, and magnetic field:  (i) transport measurements and (ii) space-resolved tunneling spectroscopy.
In the present paper, we focus on the second one and develop a theory of local density of states (LDOS) --- including both its disorder-averaged value  and fluctuations --- as measured in space-resolved tunneling experiments.

Particularly intriguing experimental findings on tunneling spectroscopy of 2D disordered superconducting systems were provided by experiments on TiN and InO films [\onlinecite{Sacepe08,Sacepe10,Sacepe11,Sherman14}]. In short, it was found that (i) the  pronounced soft gap in the tunneling spectrum survives across the superconductor-metal transition (i.e., with increasing temperature $T$ above $T_c$) and across SIT, and (ii) there are strong point-to-point fluctuations of the shape of the energy-dependence of LDOS on the superconducting side of the transition (i.e. below $T_c$). These results have been interpreted as evidence of (i) the existence of preformed Cooper pairs leading to a ``pseudogap'' in the non-superconducting states (metallic and insulating) [\onlinecite{Sacepe08,Sacepe10,Sacepe11,Sherman14}] and (ii) localization of some of Cooper pairs on the superconducting side of the transition, with the fraction of localized Cooper pairs increasing when the system approaches the SIT [\onlinecite{Sacepe11}].  Qualitatively similar features, although considerably less pronounced, were observed in experiments on NbN films
[\onlinecite{Mondal11},\onlinecite{Noat13}].  Finally, a recent work on MoC films [\onlinecite{Szabo16}] did not discover any sizeable ``pseudogap'' or spatial fluctuations effects at all; the gap observed there was related to $T_c$ by the standard formula of the Bardeen-Cooper-Schrieffer (BCS) theory.

In order to understand the experimental findings --- including features that are common for different materials as well as differences between the materials --- one clearly needs the corresponding theory. In numerical works by Ghosal, Randeria, and Trivedi, Refs.~[\onlinecite{Ghosal98},\onlinecite{Ghosal01}], a solution of  Bogoliubov-de Gennes equations for a 2D model with short-range interaction was carried out. It was found that the tunneling density of states shows a hard gap across the SIT and strong spatial fluctuations. More recently, these results were corroborated by Quantum Monte Carlo simulations [\onlinecite{Bouadim11}]. While these results are very insightful, the numerical simulations for the inherently interacting problem are limited by relatively small system sizes. This makes it difficult to explore parametric dependences of observables in a sufficiently broad range, especially since the problem is characterized by a hierarchy of relevant length and energy scales.
Such parametric dependences may be studied within analytical approaches, which are also
expected to shed more light on underlying physical mechanisms. Feigel'man, Ioffe, Kravtsov, Yuzbashyan, and Cuevas, Refs.~[\onlinecite{FeigelmanYuzbashyan2007},\onlinecite{FeigelmanCuevas2010}], studied  the  LDOS for a 3D system in the vicinity of Anderson-localization transition within a solution of the self-consistent BCS-type equation for the case of a short-range interaction.
They found that a pseudogap develops when the superconductor is built out of localized single-particle states,
and that this pseudogap increases when the system approaches the SIT.

In the present paper, we develop a theory of the LDOS of disordered superconductors which
employs the non-linear sigma-model (NLSM) formalism and the renormalization-group (RG) framework, and
goes beyond the analysis of Refs. [\onlinecite{FeigelmanYuzbashyan2007},\onlinecite{FeigelmanCuevas2010}] in several important aspects.
First, our theory takes into account mutual influence of disorder and  interaction couplings in all channels. This influence leads to the renormalization that becomes strong for systems with sufficiently strong disorder (in particular those that are not too far from SIT). Second, we consider 2D systems with short-range and Coulomb interaction on equal footing. Third, we use the same formalism to explore (i) 2D systems that would be Anderson insulators in the absence of interaction and (ii) 2D or 3D systems that undergo Anderson transition in the absence of interaction. Fourth, we evaluate both the average LDOS and its mesoscopic fluctuations (which are related to the LDOS multifractality in normal disordered systems). Fifth, when calculating the average LDOS and its moments, we take into account renormalization effects originating from all interaction channels.

On the technical side, we exploit our recent works in two complementary directions: Refs.~[\onlinecite{BurmistrovGornyiMirlin2012},\onlinecite{Burmistrov2015b}],
where the phase diagram and transport characteristics of 2D disordered systems around SIT were studied
by means of the NLSM renormalization group, on the one hand, and
Refs.~[\onlinecite{Burmistrov13},\onlinecite{Burmistrov2014},\onlinecite{Burmistrov2015a}],
where the LDOS multifractality was studied near Anderson metal-insulator transition (MIT) in a normal (i.e., not superconducting)
system with Coulomb interaction, on the other hand. Application of a unified approach to the LDOS and its fluctuations near MIT
(Refs.~[\onlinecite{Burmistrov13},\onlinecite{Burmistrov2014},\onlinecite{Burmistrov2015a}]) and in disordered superconductors (this work) turns out to be
very helpful for understanding similarities and  differences between the two cases. We will return to this issue in the end of the paper.

The outline of the paper is as follows.  In Sec. \ref{s2} we introduce the NLSM formalism
and construct operators corresponding to the moments of LDOS.
The anomalous dimensions of the moments of LDOS found within the two-loop approximation are presented in Sec. \ref{s3}.
The obtained two-loop results are used in Sec. \ref{s4} to analyze the scaling behavior of the disorder-averaged LDOS and of
the LDOS moments for the following three cases:
(i) superconducting transition in 2D system with weak short-ranged interactions;
(ii) superconducting transition in 2D system with Coulomb interaction;
(iii) superconducting transition in a system with weak short-ranged interactions which, in the absence of interactions,
is close to the Anderson transition. Our results and
conclusions are summarized in Sec. \ref{s5}.
Several appendices contain technical details on the one- and two-loop RG equations and their analysis.

\section{Formalism}
\label{s2}

\subsection{NLSM action}

We start with the description of the NLSM formalism to be used for the calculation of the local density of states and
its fluctuations near the transition to the superconducting state.
The action of the NLSM is given as a sum of the non-interacting part, $S_\sigma$ [\onlinecite{Wegner1979},\onlinecite{Efetov1980b}],
and terms $S_{\rm int}^{(\rho,\sigma,c)}$ arising from the interactions in
the particle-hole singlet and triplet, and Cooper channels [\onlinecite{Finkelstein1983},\onlinecite{Finkelstein1984}]  (see Refs. [\onlinecite{Finkelstein1990},\onlinecite{BelitzKirkpatrick1994}]
 for review):
\begin{gather}
S=S_\sigma + S_{\rm int}^{(\rho)}+S_{\rm int}^{(\sigma)}+S_{\rm int}^{(c)},
\label{eq:NLSM}
\end{gather}
where
\begin{align}
S_\sigma & = -\frac{g}{32} \int\!\! d\bm{r} \Tr (\nabla Q)^2 + 4\pi T Z_\omega \int\!\! d\bm{r} \Tr \eta  Q ,
\notag  \\
S_{\rm int}^{(\rho)}& =-\frac{\pi T}{4} \Gamma_s\! \sum_{\alpha,n} \sum_{r=0,3}
\int\!\! d\bm{r} \Tr \Bigl [I_n^\alpha t_{r0} Q\Bigr ] \Tr \Bigl [I_{-n}^\alpha t_{r0} Q\Bigr ] ,
\notag \\
S_{\rm int}^{(\sigma)}& =-\frac{\pi T}{4} \Gamma_t \sum_{\alpha,n} \sum_{r=0,3}
\int d\bm{r} \Tr \Bigl [I_n^\alpha \bm{t_{r}} Q\Bigr ]
 \Tr \Bigl [I_{-n}^\alpha \bm{t_{r}} Q\Bigr ] ,
\notag \\
S_{\rm int}^{(c)}& =-\frac{\pi T}{4}  \Gamma_c \sum_{\alpha,n} \sum_{r=1,2}  \int d\bm{r} \Tr \bigl [ t_{r0} L_n^\alpha Q \bigr ] \Tr \bigl [ t_{r0} L_n^\alpha Q \bigr ] .
\notag
\end{align}
Here we use notations from Ref. [\onlinecite{Burmistrov2015b}]. The Drude conductivity (including spin) in units $e^2/h$ is denoted as $g$. The quantities $\Gamma_s, \Gamma_t$ and $\Gamma_c$ are interaction parameters in the singlet particle-hole, triplet particle-hole, and singlet Cooper channels, respectively. The parameter $Z_\omega$ introduced by Finkelstein [\onlinecite{Finkelstein1983}] describes the renormalization of the frequency term in the action \eqref{eq:NLSM}.

The action \eqref{eq:NLSM} involves the following matrices
\begin{gather}
\Lambda_{nm}^{\alpha\beta} = \sgn n \, \delta_{nm} \delta^{\alpha\beta}t_{00},
\quad
(I_k^\gamma)_{nm}^{\alpha\beta}=\delta_{n-m,k}\delta^{\alpha\beta}\delta^{\alpha\gamma} t_{00}
,\notag \\
\eta_{nm}^{\alpha\beta}=n \, \delta_{nm}\delta^{\alpha\beta} t_{00},
\quad
(L_k^\gamma)_{nm}^{\alpha\beta}=\delta_{n+m,k}\delta^{\alpha\beta}\delta^{\alpha\gamma} t_{00} ,
\end{gather}
where $\alpha,\beta = 1,\dots, N_r$ stand for replica indices and integer numbers $n,m$ correspond to the
Matsubara fermionic energies $\varepsilon_n = \pi T (2n+1)$ and $\varepsilon_m = \pi T (2m+1)$.
The sixteen $4 \times 4$ matrices,
\begin{equation}
\label{trj}
t_{rj} = \tau_r\otimes s_j, \qquad r,j = 0,1,2,3  ,
\end{equation}
act in the spin (subsrcipt $j$) and particle-hole (subscript $r$) spaces.
The corresponding Pauli matrices are defined in a standard form as follows
\begin{equation}
\begin{split}
\tau_0 = s_0 = \begin{pmatrix}
1 & 0\\
0 & 1
\end{pmatrix}, & \qquad
\tau_1 = s_1 = \begin{pmatrix}
0 & 1\\
1 & 0
\end{pmatrix}, \\
\tau_2 = s_2= \begin{pmatrix}
0 & -i\\
i & 0
\end{pmatrix}, & \qquad \tau_3 = s_3 = \begin{pmatrix}
1 & 0\\
0 & -1
\end{pmatrix} .
\end{split}
\end{equation}
The vector $\bm{t_{r}}$ combines three $4\times 4$ matrices,
$\bm{t_{r}}=\{t_{r1},t_{r2},t_{r3}\}$.
The matrix field $Q(\bm{r})$ obeys the following constraints:
\begin{gather}
Q^2=1, \qquad \Tr Q = 0, \qquad Q^\dag = C^T Q^T C .
\label{eq:constraints}
\end{gather}
The charge conjugation matrix $C = i t_{12}$ satisfies the following relation: $C^T = -C$.
The matrix $Q$  (as well as the trace operator $\Tr$) acts in the replica, Matsubara, spin, and particle-hole spaces.

In the case of Coulomb interaction, the parameters $\Gamma_s$ and $Z_\omega$ are related to each other, $\Gamma_s=-Z_\omega$.
This relation holds in the course of the renormalization [\onlinecite{Finkelstein1983}].
This relation also reflects the symmetry of the NLSM action \eqref{eq:NLSM} under the spatially
independent rotations of the $Q$ matrix (so-called $\mathcal{F}$-invariance) [\onlinecite{Baranov1999a},\onlinecite{Burmistrov2015b}].

\subsection{Moments of the local density of states}

As usual, the local density of states $\rho(E,\bm{r})$ is expressed via the exact single-particle Green function.
Within the NLSM formalism, the disorder-averaged LDOS is determined by the operator which is linear in $Q$:
\begin{equation}
K_1(i\varepsilon_n) = \frac{\rho_0}{4} \spp \langle Q_{nn}^{\alpha\alpha}\rangle .
\label{eq:PO:TDOS}
\end{equation}
Here symbol $\spp$ denotes the trace in spin and particle-hole spaces only.
The index $\alpha$ denotes a fixed replica and $\langle \cdots \rangle$ denotes
the averaging with the NLSM action \eqref{eq:NLSM}.
The density of states at energy of the order of inverse elastic scattering time $1/\tau$ is
denoted by $\rho_0$.
We remind that $1/\tau$ plays a role of the high-energy (ultraviolet) cutoff of the NLSM theory.
The disorder-averaged LDOS $\langle \rho(E,\bm{r}) \rangle$ can be obtained after the analytic
continuation of $K_1(i\varepsilon_n)$ to the real energies, $i\varepsilon_n \to E+i0^+$.

Next, let us introduce the irreducible two-point correlation function
\begin{gather}
K_2(E,\bm{r};E^\prime, \bm{r^\prime}) = \langle \langle \rho(E, \bm{r})  \cdot \rho(E^\prime, \bm{r^\prime}) \rangle \rangle \hspace{2cm}\notag \\
\hspace{1cm}
=  \langle \rho(E, \bm{r}) \rho(E^\prime, \bm{r^\prime}) \rangle  - \langle \rho(E, \bm{r}) \rangle \langle \rho(E^\prime, \bm{r^\prime}) \rangle ,
\label{eq:K2:def:gen}
\end{gather}
which allows us to find the second moment of the LDOS.
In the NLSM approach, the correlator $K_2$ at coinciding spatial points is related to the following bilinear in $Q$ operator:
 \begin{gather}
 P_2^{\alpha_1\alpha_2}(i\varepsilon_{n},i\varepsilon_{m})
 = \langle\langle  \spp Q_{nn}^{\alpha_1\alpha_1}(\bm{r}) \cdot \spp Q_{mm}^{\alpha_2\alpha_2}(\bm{r}) \rangle \rangle \notag  \\
 - 2 \langle \spp \bigl [Q_{nm}^{\alpha_1\alpha_2}(\bm{r}) Q_{mn}^{\alpha_2\alpha_1}(\bm{r}) \bigr ] \rangle
  , \quad \alpha_1 \neq \alpha_2 .
 \label{eqP2corr}
 \end{gather}
The correlation function $K_2(E,\bm{r};E^\prime, \bm{r})$ defined for real energies
can be obtain from the following Matsubara counterpart
 \begin{equation}
K_2 = \frac{\rho_0^2}{32}\re \Bigl [ P_2^{\alpha_1\alpha_2}(i\varepsilon_{n_1},i\varepsilon_{n_3}) - P_2^{\alpha_1\alpha_2}(i\varepsilon_{n_1},i\varepsilon_{n_2}) \Bigl ]
 \label{eqK2def0}
 \end{equation}
after analytic continuation:
$\varepsilon_{n_1} \to E+i0^+$, $\varepsilon_{n_3} \to E^\prime+i0^+$,
and $\varepsilon_{n_2} \to E^\prime-i0^+$.
We use the convention that $n_1, n_3, n_5, \dots \geqslant 0$ and $n_2, n_4, n_6 < 0$.

The following comments are in order here.
\begin{itemize}
\item[(i)] The condition that replica
indices $\alpha_1$ and $\alpha_2$ are nonequal in Eq. \eqref{eqK2def0} stems
from the fact that the two-point correlation function $K_2$ measures
\textit{mesoscopic fluctuations} of the LDOS.
This forbids interaction lines between two fermionic loops corresponding
to the LDOS in the diagrammatic approach.
\item[(ii)] The bilinear-in-$Q$ operator \eqref{eqK2def0} is the eigenoperator under the
action of the RG  (below we shall explicitly prove this statement by means of the two-loop calculations).
\item[(iii)] Disorder-averaged higher moments of the LDOS can be expressed in terms
of higher-order irreducible correlation functions of the $Q$-field similarly to Eq. \eqref{eqK2def0}.
Explicit examples of operators corresponding to the third and forth moments of the LDOS can
be found in Ref. [\onlinecite{Burmistrov2015a}].
The corresponding operators are also eigenoperators of the renormalization group.
\end{itemize}

\section{Renormalization group for LDOS}
\label{s3}

In this section we outline the RG formalism in the context of the calculation of the moments of
the distribution of the local DOS as derived from the NLSM.

\subsection{Perturbative expansion}

To resolve the nonlinear constraint  \eqref{eq:constraints} we adopt the square-root parametrization:
\begin{gather}
Q = W +\Lambda \sqrt{1-W^2}\ , \qquad W= \begin{pmatrix}
0 & w\\
\bar{w} & 0
\end{pmatrix} .
\label{eq:Q-W}
\end{gather}
We use the following notations: $W_{n_1n_2} = w_{n_1n_2}$ and $W_{n_2n_1} = \bar{w}_{n_2n_1}$ with $n_1\geqslant 0$ and $n_2< 0$.
As a consequence of the charge-conjugation constraint \eqref{eq:constraints},
the blocks $w$ and $\bar{w}$ obey the following relations:
\begin{gather}
\bar{w} = -C w^T C,\qquad w = - C w^* C .
\end{gather}
These relations imply that elements
$(w^{\alpha\beta}_{n_1n_2})_{rj}$  in the expansion $w^{\alpha\beta}_{n_1n_2}= \sum_{rj} (w^{\alpha\beta}_{n_1n_2})_{rj} t_{rj}$
are real or purely imaginary.
The perturbative (in $1/g$) analysis of the NLSM action \eqref{eq:NLSM} is performed by expanding
the action in powers of $W$.

From the expansion the NLSM action \eqref{eq:NLSM} to the second order in $W$,
we find the bare propagators (see Ref. [\onlinecite{BelitzKirkpatrick1994}]).
The propagators in the particle-hole channel (diffusons) read ($r=0,3$ and $j=0,1,2,3$)
\begin{gather}
\Bigl \langle [w_{rj}(\bm{p})]^{\alpha_1\beta_1}_{n_1n_2} [\bar{w}_{rj}(-\bm{p})]^{\beta_2\alpha_2}_{n_4n_3} \Bigr \rangle =  \frac{2}{g} \delta^{\alpha_1\alpha_2} \delta^{\beta_1\beta_2}\delta_{n_{12},n_{34}}\notag \\
\times  \mathcal{D}_p(i\Omega_{12}^\varepsilon)\Bigl [\delta_{n_1n_3} - \frac{32 \pi T \Gamma_j}{g}\delta^{\alpha_1\beta_1}  \mathcal{D}_p^{(j)}(i\Omega_{12}^\varepsilon) \Bigr ] ,
\label{eq:prop:PH}
\end{gather}
where $n_{12}=n_1-n_2$ and $\Omega_{12}^\varepsilon = \varepsilon_{n_1}-\varepsilon_{n_2}$.
The standard diffusive propagator is given by
\begin{equation}
\mathcal{D}^{-1}_p(i\omega_n) =p^2+{16 Z_\omega |\omega_n|}/{g},
\label{eq:prop:Free}
\end{equation}
with $\omega_n = 2\pi T n$.
The diffusons renormalized by interaction in the particle-hole channels read
\begin{equation}
[\mathcal{D}^{(j)}_p(i\omega_n)]^{-1}  =  p^2+{16 (Z_\omega+\Gamma_j) |\omega_n|}/{g} .
 \label{eq:prop:Int}
\end{equation}
The propagators in the particle-particle channel (cooperons) can be written as ($r=1,2$ and $j=0,1,2,3$)
\begin{gather}
\Bigl \langle [w_{rj}(\bm{q})]^{\alpha_1\beta_1}_{n_1n_2} [\bar{w}_{rj}(-\bm{q})]^{\beta_2\alpha_2}_{n_4n_3} \Bigr \rangle =  \frac{2}{g} \delta^{\alpha_1\alpha_2} \delta^{\beta_1\beta_2}\delta_{n_{14},n_{32}}\notag \\
\times  \mathcal{C}_q(i\Omega_{12}^\varepsilon)\Bigl [\delta_{n_1n_3} - \frac{64 \pi T Z_\omega}{g}\delta^{\alpha_1\beta_1} \delta_{j0} \mathcal{C}_q(i\Omega_{34}^\varepsilon)  \mathcal{L}_q(i\mathcal{E}_{12}) \Bigr ] ,
\label{eq:prop:PPS}
\end{gather}
where $\mathcal{E}_{12} = \varepsilon_{n_1}+\varepsilon_{n_2}$ and
$\mathcal{C}_q(i\omega_n) \equiv \mathcal{D}_q(i\omega_n)$.
The propagator $\mathcal{L}$ stands for the standard
superconducting-fluctuation propagator:
\begin{align}
\mathcal{L}^{-1}_q(i\omega_n) = \gamma_c^{-1}
+ \ln \frac{\Lambda_U}{4\pi T} & - \psi\left (\frac{D q^2 + |\omega_n|}{4\pi T}+\frac{1}{2} \right ) \notag \\
& + \psi\left (\frac{1}{2} \right ) .
\label{eq:prop:fp}
\end{align}
Here we have introduced $\gamma_c= \Gamma_c/Z_\omega$.
The quantity $D = g/(16 Z_\omega)$ is the diffusion coefficient, $\Lambda_{U} \sim 1/\tau$
is the ultraviolet energy scale, and $\psi(z)$ denotes the digamma function.
We note that the fluctuation propagator \eqref{eq:prop:fp}
is written under the assumption that
the infrared energy scale of the theory is determined by temperature.

For the purpose of regularization in the infrared, we add the extra term
\begin{equation}
S \to S + \frac{g h^2}{8} \int d \bm{r}\Tr \Lambda Q
\label{SsGenFull}
\end{equation}
to the NLSM action \eqref{eq:NLSM}.
The presence of this term in the action results, in particular, in the substitution
of $p^2 +h^2$ for $p^2$ in the propagators \eqref{eq:prop:Free},  \eqref{eq:prop:Int}, and \eqref{eq:prop:fp}.

\subsection{The disorder-averaged LDOS}

For the sake of completeness, and in order to set notations, we remind the reader
the result of one-loop renormalization of the disorder-averaged LDOS.
Expanding the matrix $Q$ to the second order in $W$, we derive from Eq. \eqref{eq:PO:TDOS}
the following expression:
\begin{equation}
\frac{K_1(i\varepsilon_{n_1})}{\rho_0}
= 1 - \frac{1}{8} \spp \sum_{n_2,\beta}
\langle w_{n_1n_2}^{\alpha\beta}(\bm{r}) \bar{w}_{n_2n_1}^{\beta\alpha}(\bm{r})  \rangle  .
\end{equation}
Next, using Eqs. \eqref{eq:prop:PH} and \eqref{eq:prop:PPS}, we find
\begin{gather}
\frac{\rho(i\varepsilon_{n_1})}{\rho_0}
= 1 + \frac{64\pi T}{g^2} \int_q \sum_{\omega_n>\varepsilon_{n_1}}  \sum_{j=0}^3
\Gamma_j \mathcal{D}_q(i\omega_n) \mathcal{D}^{(j)}_q(i\omega_n) \notag \\
+ \frac{128\pi T Z_\omega}{g^2}
\int_q \sum_{\omega_n<\varepsilon_{n_1}}\mathcal{C}^2_q(2i\varepsilon_{n_1}-i\omega_n)
\mathcal{L}_q(i\omega_n).
\label{app:eq_P1}
\end{gather}
Finally,  performing the analytic continuation to real frequencies,
$i \varepsilon_{n_1} \to E + i 0^+$, we obtain
that the disorder-averaged LDOS can be written as
\begin{equation}
\langle \rho(E) \rangle = \rho_0 [Z(E)]^{1/2},
\label{rhoZ}
\end{equation}
where  the renormalization factor $Z(E)$ is given by (see Refs. [\onlinecite{AAbook},\onlinecite{KLbook}] for a review):
\begin{gather}
Z(E)   = 1   +  \frac{32}{g^2} \im  \sum_{j=0}^3 \Gamma_j
\int_{q,\omega}  \mathcal{F}_{\omega-E}   \mathcal{D}^R_q(\omega) \mathcal{D}^{(j) R}_q(\omega)
 \notag \\
+  \frac{64 Z_\omega}{g^2} \im \int_{q,\omega} \mathcal{C}^{R2}_q(2E-\omega) \bigl [
  \mathcal{L}^K_q(\omega)+\mathcal{F}_{E-\omega} \mathcal{L}^R_q(\omega)
 \bigr ] .
\label{rhores}
\end{gather}
Here $\mathcal{D}^R_q(\omega)$, $\mathcal{C}^R_q(\omega)$, $\mathcal{D}^{(j) R}_q(\omega)$, and $\mathcal{L}^R_q(\omega)$
are retarded propagators obtained from the corresponding Matsubara propagators.
The Keldysh part of the fluctuation propagator $\mathcal{L}^K_q(\omega)$ is related to the
retarded one via the bosonic distribution function $\mathcal{B}_\omega = \coth(\omega/2T)$ as follows:
$\mathcal{L}^K_q(\omega) = 2i \mathcal{B}_\omega \im \mathcal{L}^R_q(\omega)$.
The fermionic distribution function is denoted as $\mathcal{F}_\omega = \tanh(\omega/2T)$.
We adopt the following short-hand notation:
\begin{equation}
\int_{q,\omega} \equiv \int \frac{d^d\bm{q}}{(2\pi)^d} \int_{-\infty}^\infty d\omega \, .
\end{equation}
We note that the definition (\ref{rhoZ}) of $Z$ coincides with the definition of the field
renormalization constant in Ref. [\onlinecite{BelitzKirkpatrick1994}] and the
wavefunction renormalization constant in Ref. [\onlinecite{Castellani1984}].
We stress that $Z$ differs from the Finkel'stein's frequency renormalization
factor $Z_\omega$ and should not be confused with the latter.

To derive the RG equation for $Z$, we set temperature $T$ and energy $E$ to
zero and study the dependence of $Z$ on the infrared regulator $h^2$.
Then in $d=2+\epsilon$ dimensions we obtain
 \begin{equation}
Z = 1 - \bigl [\ln(1+\gamma_s)+3\ln(1+\gamma_t)+2\gamma_c\bigr ] \frac{h^\epsilon t}{\epsilon} +O(\epsilon) .
\label{eqAvDOS}
\end{equation}
Here $\gamma_{s,t}=\Gamma_{s,t}/Z_\omega$ are dimensionless interaction amplitudes and $t = 8 \Omega_d/g$ denotes
dimensionless resistivity, with $\Omega_d= S_d/[2(2\pi)^d]$ and $S_d=2\pi^{d/2}/\Gamma(d/2)$ being
the area of a $d$-dimensional sphere.
As usual, Eq. \eqref{eqAvDOS} determines the anomalous dimension $\zeta$ of the disorder-averaged LDOS.
Using the minimal subtraction scheme (see e.g., Ref. [\onlinecite{Amit-book}]), we obtain  in the one-loop approximation [\onlinecite{Altshuler1979},\onlinecite{Finkelstein1984},\onlinecite{Castellani1984}]:
\begin{equation}
- \frac{d \ln Z}{d y} = 2 \zeta = - \bigl [\ln(1+\gamma_s)+3\ln(1+\gamma_t)+2\gamma_c\bigr ] t + O(t^2) ,
\label{eq:ad:Z}
\end{equation}
where $y=\ln 1/h$ is the running RG length scale.
We note that a more accurate treatment of the term
with the fluctuation propagator in Eq. \eqref{rhores} (see Appendix A of Ref. [\onlinecite{Burmistrov2015b}])
results in exactly the same RG equation as Eq. \eqref{eq:ad:Z}.
Thus, this one-loop RG equation is formally exact in all three interaction
couplings $\gamma_{s}$, $\gamma_t$, and $\gamma_c$. In the case of fully broken spin-rotational
symmetry, Eq. \eqref{eq:ad:Z} holds with the contribution $3\ln(1+\gamma_t)$ of the triplet
particle-hole channel in the right hand side being omitted.

\subsection{The second moment of the LDOS}

Now we consider the renormalization of the second moment of the LDOS.
We restrict our consideration by one- and two-loop orders in $t$.

\subsubsection{One-loop results}

In the one-loop approximation one obtains
 \begin{gather}
[P_2^{\alpha_1\alpha_2}]^{(1)}(i\varepsilon_{n_1},i\varepsilon_{n_3}) = 0,
\end{gather}
and
\begin{gather}
[P_2^{\alpha_1\alpha_2}]^{(1)}(i\varepsilon_{n_1},i\varepsilon_{n_2}) = - 2 \spp \langle w^{\alpha_1\alpha_2}_{n_1n_2}(\bm{r}) \bar{w}^{\alpha_2\alpha_1}_{n_2n_1}(\bm{r}) \rangle \notag \\ = - \frac{128}{g} \int_q
\Bigl [ \mathcal{D}_q(i\Omega^{\varepsilon}_{12})+\mathcal{C}_q(i\Omega^{\varepsilon}_{12})\Bigr]
.  \label{eq18P1}
\end{gather}
Hence, we find
\begin{equation}
K_2^{(1)}(E,\bm{r};E^\prime, \bm{r})  = \rho_0^2\,\frac{4 }{g} \re \int_q \Bigl [  \mathcal{D}^R_q(\Omega) +
 \mathcal{C}^R_q(\Omega)\Bigr ]  , \label{eq1loopK2_2}
\end{equation}
where $\Omega=E-E^\prime$.
Setting $\Omega=0$ and using $h^2$ as the infrared regulator, we get
\begin{equation}
K_2^{(1)}  =  - \rho_0^2 \frac{2 h^\epsilon t}{\epsilon} +O(\epsilon) . \label{eq1loopK2_3}
\end{equation}

\subsubsection{Two-loop results}

Details of the calculation of the two-loop contribution to the correlation function $K_2$ are presented in Appendix \ref{Sec:App:2loop}.
Using Eqs. \eqref{eq2loopK2_P2++2} and \eqref{eq2loopK2_P2+-2} of Appendix \ref{Sec:App:2loop},
we find the following two-loop contribution to the irreducible two-point correlation function:
\begin{align}
K_2^{(2)}  & = \rho_0^2 \frac{t^2\,h^{2\epsilon}}{\epsilon^2} \Biggl \{ 1+2 \Bigl (1 +\frac{\epsilon}{2}\Bigr ) +
\Bigl (3 +\epsilon\Bigr ) \gamma_c +\sum_{j=0}^3 \Bigl [   f(\gamma_j)  \notag \\
& +2 \ln(1+\gamma_j)  + \frac{\epsilon}{2} \bigl [ \ln(1+\gamma_j) + 2 f(\gamma_j) - c(\gamma_j)
\bigr ]\Bigr ] \Biggr \} ,
\label{eqK2_2}
\end{align}
Here the function
$\liq(x) = \sum_{k=1}^\infty x^k/k^2$ denotes the polylogarithm,
\begin{gather}
c(\gamma) = 2 +\frac{2+\gamma}{\gamma} \liq(-\gamma) + \frac{1+\gamma}{2\gamma} \ln^2(1+\gamma),
\label{eq:def:c}
\end{gather}
and
\begin{equation}
\label{fx}
f(x) = 1 - (1+1/x)\ln(1+x) .
\end{equation}

\subsubsection{Anomalous dimension of the second moment of LDOS}

Above we have derived the dependence of $K_2$ on the momentum scale $h$ within
the two-loop approximation. However, $h$ itself acquires renormalization [\onlinecite{Baranov1999}].
The renormalized momentum scale  $h^\prime$ is defined as follows
\begin{equation}
g^\prime h^{\prime 2}  \Tr \Lambda^2 = g h^2 \langle \Tr \Lambda Q \rangle ,
\label{eq:ren:h}
\end{equation}
where $g^\prime$ denotes the renormalized conductivity at the momentum
scale $h^\prime$.  As a consequence of Eq. \eqref{eq:ren:h}, $h^\prime$ satisfies the following
relation: $g^\prime h^{\prime 2} = g h^2 Z^{1/2}$.
Using Eq. \eqref{eqAvDOS} and the one-loop result for the conductivity [\onlinecite{Altshuler1979b},\onlinecite{Altshuler1980},\onlinecite{Finkelstein1983},\onlinecite{Castellani1984}]
\begin{gather}
g^\prime = g \Bigl [1+\frac{a_1 t\, h^\epsilon}{\epsilon} +O(\epsilon)\Bigr ], \quad
 a_1 =1+\sum_{j=0}^3 f(\gamma_j) - \gamma_c ,
\label{eqS1}
\end{gather}
we find
\begin{equation}
h^\prime = h \Biggl \{  1 - \frac{t\,  h^\epsilon}{2\epsilon}
\Bigl [ 1+ \sum_{j=0}^3\bigl [ f(\gamma_j)+\frac{1}{2}\ln(1+\gamma_j) \bigr ] \Bigr ] \Biggr \}  .
\label{eqhren}
\end{equation}
We note that within the one-loop approximation, there is no contribution to Eq.~(\ref{eqhren}) due to interaction in the Cooper channel.

By using Eqs. \eqref{eq1loopK2_3}, \eqref{eqK2_2}, and \eqref{eqhren},
we can rewrite the second moment of the LDOS in terms of the renormalized momentum
scale $h^\prime$  and the renormalization factor $Z$ as follows:
\begin{equation}
\langle \rho^2 \rangle = Z \rho_0^2 +K_2 = \rho_0^2 Z m^\prime_2,
\label{eq:2dm}
\end{equation}
where
\begin{equation}
m^\prime_2 =  m_2 \Bigl [ 1+ \frac{b^{(2)}_1 t \, h^{\prime \epsilon}}{\epsilon}+ \frac{t^2 h^{\prime 2\epsilon}}{\epsilon^2} \Bigl (b^{(2)}_2+\epsilon b^{(2)}_3\Bigr ) \Bigr ] .
\label{eqM2def}
\end{equation}
Here we omit the terms that are finite in the limit $\epsilon\to 0$.
The bare value of $m^\prime_2$ is unity, $m_2=1$, and
\begin{gather}
b^{(2)}_1 = -2, \quad b^{(2)}_2 = 3+\sum_{j=0}^3 f(\gamma_j) -  \gamma_c, \notag \\
 b^{(2)}_3 = - \frac{1}{2} \sum_{j=0}^3 c(\gamma_j) + \gamma_c.
\end{gather}
Next, we introduce a dimensionless quantity
$\bar{t} = t^\prime h^{\prime \epsilon}$.
With the help of  Eqs \eqref{eqS1} and \eqref{eqM2def},
we express $t$, $\gamma_j$ and $m_2$  as follows:
\begin{gather}
t =  (h^\prime)^{-\epsilon} \bar{t} Z_t(\bar{t},\gamma^\prime_s,\gamma^\prime_t),\qquad \gamma_j = \gamma_j^\prime Z_j(\bar{t},\gamma^\prime_s,\gamma_t^\prime), \notag \\
 m_2= m_2^\prime
Z_{m_2}(\bar{t},\gamma^\prime_s,\gamma^\prime_t) .
\end{gather}
The interaction parameters $\gamma_j$ are renormalized
at the one-loop level [\onlinecite{Finkelstein1983}].
However this  does not affect the two-loop result
for the anomalous dimension of $m^\prime_2$
since $b^{(2)}_1$ is independent of $\gamma_j$.
To the lowest orders in $\bar{t}$, the renormalization parameters become
\begin{equation}
Z_t= 1 + \frac{a_1}{\epsilon}\bar{t},
\label{eqZZZ2}
\end{equation}
and
\begin{equation}
Z_{m_2}^{-1} =  1 + \frac{b^{(2)}_1}{\epsilon}\bar{t} +
\frac{\bar{t}^2}{\epsilon^2} \Bigl [ b^{(2)}_2 + b^{(2)}_1 a_1 + \epsilon b^{(2)}_3 \Bigr ] .
\label{eqZm2}
\end{equation}
The anomalous dimension of $m_2^\prime$ is derived from standard conditions
that $m_2$ (as well as $t$ and $\gamma_j$) does not depend on the momentum
scale $h^\prime$. In this way, we obtain the following two-loop result for the anomalous
dimension $\zeta_2(t,\gamma_s,\gamma_t,\gamma_c)$ of $m_2^\prime$:
\begin{equation}
-\frac{d \ln m_2}{d y} = \zeta_2 =  -2 t - [c(\gamma_s)+3c(\gamma_t)-2\gamma_c] t^2 + O(t^3) .
\label{eqm2RG}
\end{equation}
Here we omit `prime' and `bar' signs for brevity.
We remind the reader that the function $c(\gamma)$ is defined in Eq. \eqref{eq:def:c}.
We note that $c(0) = 0$ as it is known for free electrons [\onlinecite{Wegner1979}].
In the case of Coulomb interaction, one has $c(-1) = 2 -\pi^2/6\approx 0.36$.
The interaction affects the anomalous dimension at the two-loop order only.

We emphasize that coefficients $a_1$, $b^{(2)}_1$ and $b^{(2)}_2$ satisfy the relation
\begin{equation}
b^{(2)}_2 = b^{(2)}_1(b^{(2)}_1-a_1)/2.
\label{eq:rel:rg}
\end{equation}
This guaranties the absence in Eq.~\eqref{eqm2RG} of terms divergent in
the limit $\epsilon\to 0$, i.e. the renormalizability of $m_2$.
In addition, this relation proves that the operator corresponding to $K_2$ is the
RG eigenoperator. Indeed, if the operator corresponding to $K_2$ is the linear combination
of several eigenoperators, the relation \eqref{eq:rel:rg} would imply non-linear system
of equations which has no non-trivial solutions in general.

Combining the above results, the second moment of the LDOS can be written as
\begin{equation}
\langle \rho^2 \rangle = \langle \rho \rangle^2\, m_2 .
\label{eqK3}
\end{equation}
The scaling behavior of $\langle \rho \rangle$ and $m_2$ are
governed by the anomalous dimensions~ \eqref{eq:ad:Z} and \eqref{eqm2RG}, respectively.

\subsection{The $q$-th moment of the LDOS}

In this section, we generalize the results obtained in the previous section for the $q$-th moment of the LDOS.
The important observation is that the irreducible $q$-th moment of the LDOS,
$K_q = \bigl \langle \bigl ( \rho - \langle \rho \rangle \bigr )^q \bigr \rangle$,
involves connected contributions from averages of the number $q$ of matrices $Q$.
Therefore, $K_q$ has no one- and two-loop contributions for $q\geqslant 5$.
Consequently, as in the case of noninteracting electrons [\onlinecite{Wegner1986,Wegner1987a,Wegner1987b}],
the anomalous dimension for the $q$-th moment of the LDOS becomes proportional to the factor $q(1-q)$ within
one- and two-loop approximation (see details in Ref. [\onlinecite{Burmistrov2015a}]).
Thus, we find
\begin{equation}
\left \langle \rho^q\right \rangle =\langle \rho\rangle^q m_q  ,
\label{eqKmq}
\end{equation}
where the behavior of $m_q$ is determined by the following RG equation:
\begin{equation}
-\frac{d \ln m_q}{d y} = \zeta_q =\frac{q(1-q)}{2} \Bigl \lbrace 2 t +  \bigl [c(\gamma_s)+3c(\gamma_t)  -2\gamma_c\bigr ] t^2 \Bigr \rbrace  .
\label{eqmqRG}
\end{equation}
Here the function $c(\gamma)$ is defined in Eq. \eqref{eq:def:c}.
We note that for the special cases $q=3$ and $q=4$, one can demonstrate that Eq. \eqref{eqmqRG}
holds as well (see Appendix C of Ref. [\onlinecite{Burmistrov2015a}]).

As we have already mentioned above, the two-loop contributions to $[P_2^{\alpha_1\alpha_2}]^{RA}$
can be interpreted as the renormalization of diffusions and cooperons involved
in the one-loop term $[P_1^{\alpha_1\alpha_2}]^{RA}$ (see Appendix \ref{app:sec:2}).
Therefore, within the two-loop approximation, the corrections due to
fluctuating Cooper pairs to $m_q$ comes from the term $[P_1^{\alpha_1\alpha_2}]^{RR}$ only.
In the one-loop approximation the fluctuation corrections to the $q$-th moment of the LDOS are fully determined by those in the average
LDOS via the factor $\langle \rho\rangle^q$.

In the absence of spin-rotational symmetry, the anomalous dimension  $\zeta_q$
can be obtained from Eq. \eqref{eqmqRG} as follows:
(i) one omits the contribution $3c(\gamma_t)$ of the triplet particle-hole channel
and (ii) one multiples the right hand side of Eq. \eqref{eqmqRG} by the factor $1/4$ (see Ref. [\onlinecite{Burmistrov2015a}]).
Thus, in the case of broken spin rotational symmetry, Eq. \eqref{eqmqRG} takes the following form:
\begin{equation}
-\frac{d \ln m_q}{d y} = \zeta_q =\frac{q(1-q)}{2} \left\lbrace \frac{t}{2} +  \bigl [c(\gamma_s)  -2\gamma_c\bigr ] \frac{t^2}{4} \right\rbrace  .
\label{eqmqRG-SO}
\end{equation}

\section{Scaling analysis}
\label{s4}

\subsection{Weak coupling RG equations in 2D}

Recently [\onlinecite{Burmistrov2015b}], the full set of one-loop RG equations describing
the renormalization of  resistivity and interactions has been derived by means of
the background field renormalization of the NLSM \eqref{eq:NLSM}:
\begin{align}
\frac{d t}{dy} & = t^2 \Bigl [ \frac{\mathcal{N}-1}{2} + f(\gamma_s)+\mathcal{N} f(\gamma_t)- \gamma_c \Bigr ] , \label{eq:rg:final:t}\\
\frac{d\gamma_s}{dy}  & = - \frac{t}{2} (1+\gamma_s)\bigl ( \gamma_s+\mathcal{N}\gamma_t+2\gamma_c+4\gamma_c^2\bigr ), \label{eq:rg:final:gs} \\
\frac{d\gamma_t}{dy}  & = - \frac{t}{2} (1+\gamma_t) \Bigl [\gamma_s-(\mathcal{N}-2)\gamma_t-2\gamma_c \notag\\
& \hspace{2cm}
-4\gamma_c\gamma_t+4\gamma_c^2 \Bigr ], \label{eq:rg:final:gt} \\
\frac{d\gamma_c}{dy} & = - 2\gamma_c^2 - \frac{t}{2} \Bigl [ (1+\gamma_c)(\gamma_s-\mathcal{N}\gamma_t) - 2\gamma_c^2+4\gamma_c^3\notag \\
 & \hspace{2cm} + 2\mathcal{N}\gamma_c \Bigl (\gamma_t-\ln(1+\gamma_t)\Bigr )\Bigr ] ,
\label{eq:rg:final:gc}
\\
\frac{d\ln Z_\omega}{dy} & = \frac{t}{2} \Bigl (\gamma_s+\mathcal{N}\gamma_t+2\gamma_c +4\gamma_c^2\Bigr ) .\label{eq:rg:final:Z}
\end{align}
Here  $\mathcal{N}$ stands for the number of triplet diffusive modes.
In the case of preserved spin-rotational symmetry, all three triplet diffusons contribute to the RG equations, $\mathcal{N}=3$.
If spin-rotational symmetry is  broken, all triplet modes are suppressed at long lengthscales, $\mathcal{N}=0$.
In addition, in this case RG equation for $\gamma_t$ should be ignored.

The above RG equations are derived in the lowest order in $t\ll 1$.
Extending the previous result [\onlinecite{Finkelstein1984}], in this order
they contains \textit{all} contributions due to the interaction in the Cooper channel, $\gamma_c$.
Comparison of the disorder-independent and disorder-induced terms in the right hand side
of Eq. \eqref{eq:rg:final:gc} demonstrates that one-loop RG equations can be used towards the
superconducting instability upto the length scale $L_X$ at which $|\gamma_c(L_X)| t(L_X) \sim 1$
[\onlinecite{Burmistrov2015b}].
We note that similar conclusion follows from comparison of one and two-loop contributions
in Eqs. \eqref{eqmqRG} and \eqref{eqmqRG-SO}.

\subsection{Weak short-ranged interactions in 2D}
\label{s4b}

We start our analysis of the fluctuations of LDOS with the case of weak short-ranged
interactions, $|\gamma_{s,t,c}| \ll 1$, in two dimensions. The phase diagram and transport properties for this case
have been discussed in details in Refs. [\onlinecite{BurmistrovGornyiMirlin2012},\onlinecite{Burmistrov2015b}].
The existence of a large region of the superconducting phase with transition
temperature higher than standard expression of BCS theory has been predicted.

For the sake of convenience, we briefly remind the reader
the main steps of analysis of Ref. [\onlinecite{BurmistrovGornyiMirlin2012}].
Let us focus on the case of preserved spin-rotational symmetry.
For $|\gamma_{s,t,c}| \ll 1$, the set of RG equations \eqref{eq:rg:final:t} - \eqref{eq:rg:final:gc} can be simplified to
\begin{equation}
dt / dy = t^2 , \,
\label{eq:1:e1}
\end{equation}
\begin{equation}
\label{eq:1:e2}
\frac{d}{dy} \left(\!\! \begin{array}{c} \gamma_s \\ \gamma_t \\ \gamma_c
\end{array}\!\! \right) = -\frac{t}{2} \left( \!\!\begin{array}{ccc}
1 & 3 & 2 \\
1 & -1 & -2 \\
1 & -3 & 0\\
\end{array}\!\! \right)
\left( \!\!\begin{array}{c} \gamma_s \\ \gamma_t \\ \gamma_c
\end{array}\!\! \right) -
\left(\!\! \begin{array}{c} 0 \\ 0 \\ 2\gamma_c^2
\end{array}\!\! \right) \, .
\end{equation}
Equation~\eqref{eq:1:e1} yields usual weak localization behavior:
\begin{equation}
\label{eq:1:e4}
t^{-1}(y) = t_0^{-1} - y \, .
\end{equation}
Here $t_0 \ll 1$ denotes the bare value of resistance. In the absence of interactions,
Eq. \eqref{eq:1:e4} suggests that a strong Anderson insulator emerges at the scale $y_{I} = t_0^{-1}$.
In the case of not too weak interactions (or sufficiently weak disorder),
$t_0\ll |\gamma_{s,0}|, |\gamma_{t,0}|, |\gamma_{c,0}|$,
Eqs. \eqref{eq:1:e2} reduce to the standard RG equation for the BCS instability
in the clean case. Then the superconducting transition occurs at the scale
of the order of $y_{\rm BCS} = 1/(2|\gamma_{c,0}|)$.

For the case of disorder which is strong compared to the
interaction, $t_0 \gg |\gamma_{s,0}|, |\gamma_{t,0}|, |\gamma_{c,0}|$, the renormalization proceeds in two steps.
At the first step, we can neglect the $-2\gamma_c^2$ term.
This brings us to a linear system of equations.
The corresponding $3\times 3$ matrix has two eigenvalues:
\begin{equation}
\lambda = 2t,\quad \lambda^\prime = -t,
\end{equation}
where $\lambda^\prime$ is doubly degenerate.
We emphasize that the eigenvalue $\lambda$ coincides with the one-loop result for $\zeta_2$.
This occurs since the interactions in the NLSM action \eqref{eq:NLSM} are described by the
operators bilinear in $Q$.
Thus, three couplings are tend to the eigenvector $\{-1,1,1\}$ corresponding to the positive eigenvalue $\lambda$.
This eigenvector  parametrizes the so-called BCS line
\begin{equation}
-\gamma_s = \gamma_t = \gamma_c \equiv \gamma.
\label{BCSline}
\end{equation}
It is this relation between the couplings that one obtains starting from a standard
BCS Hamiltonian with the attraction only.
Projecting Eqs. \eqref{eq:1:e2} onto the BCS line, we get
\begin{equation}
\label{eq:1:e5}
d\gamma /dy = 2 t \gamma - 2\gamma^2/3 \,.
\end{equation}
We note that the RG flow in directions perpendicular to the eigenvector $\{-1,1,1\}$ does
not affect the results in any essential way (see Ref. [\onlinecite{BurmistrovGornyiMirlin2012}] for details).

We are interested in the case when the dominant bare
interaction is the Cooper attraction $\gamma_{c,0}<0$ such that the initial value
$\gamma_{0}= (-\gamma_{s,0} + 3\gamma_{t,0} + 2\gamma_{c,0})/6 <0$.
Equation~\eqref{eq:1:e5} describes two distinct scenarios.
For $|\gamma_{0}| \ll t_0^2$, the resistance $t$ becomes of order unity when $|\gamma|$ is still small.
This means that, with further increase of the length scale, the system flows toward a strong Anderson
insulator.

In the opposite case, $|\gamma_{0}| \gg t_0^2$,  $|\gamma|$ increases under
the RG transformation due to the first term in the r.h.s. of Eq.~\eqref{eq:1:e5}.
The attractive interaction overtakes $t$ and reaches unity at the scale
$y_c \approx t_0^{-1} [1- 2|\gamma_0|/(3 t_0)]$.
We note that
\begin{equation}
t(y_c)\equiv t_c = 3t_0^2/(2|\gamma_0|) \ll 1,
\label{eq:def:tst}
\end{equation}
i.e. strong attraction is arisen in the region of good metal.
In this situation, we expect that with further increase of the length scale,
the RG flow develops a superconducting instability (due to standard BCS-type mechanism).
The temperature $T_c$ of this superconducting transition can be estimated as
\begin{equation}
\label{eq:1:e6}
T_c \sim \frac{1}{\tau} e^{-2y_c}
\sim \frac{1}{\tau} \exp\left (-\frac{2}{t_0} + \frac{2}{t_c}\right ) \, .
\end{equation}

In the considered regime of sufficiently strong disorder,
$t_0 \gg |\gamma_{s,0}|, |\gamma_{t,0}|, |\gamma_{c,0}|$, the temperature
given by Eq.~(\ref{eq:1:e6}) is much higher than the clean BCS value
\begin{equation}
\label{eq:1:e7}
T_c^{\rm BCS} \sim \tau^{-1} e^{-2 y_{\rm BCS}} \sim \tau^{-1} e^{- 1/{|\gamma_{c,0}|}}  .
\end{equation}
When $t_0$ becomes smaller than $|\gamma_0|$, Eq.~\eqref{eq:1:e6} crosses over into Eq. \eqref{eq:1:e7}.
Therefore, $T_c$ shows a non-monotonous dependence on the disorder strength and gets strongly
enhanced in the intermediate range of resistivity, $|\gamma_0|\ll t_0\ll \sqrt{|\gamma_0|}$.
For a given interaction strength $\gamma_0=(-\gamma_{s,0} + 3\gamma_{t,0} + 2\gamma_{c,0})/6<0$,
the superconducting critical temperature $T_c$ is the largest
when the system approaches the superconductor-insulator transition.
The latter takes place at $t_c\sim 1$, i.e. $t_0 \sim \sqrt{|\gamma_{0}|}$.
It is worth noting that at the superconductor-insulator transition the critical temperature on the superconducting side
is given by $T_c \sim \delta_\xi = \tau^{-1} \xi^{-2}$, where  $\xi \sim l \exp(-2/t_0)$ is the localization
length defined such as $t(\xi) \sim 1$.

The above analysis [\onlinecite{BurmistrovGornyiMirlin2012}] for $|\gamma_{s,t,c}| \ll 1$
was extended in Ref. [\onlinecite{Burmistrov2015b}] to the case of strong interaction couplings.
The numerical integration of the full RG equations \eqref{eq:rg:final:t} - \eqref{eq:rg:final:gc} 
provided the phase diagram of the SIT. In particular, it was found that, when the system is initially 
close to the BCS line, the enhancement of $T_c$ occurs for ${|\gamma_{c,0}|\alt\!0.2}$.

Now let us return to the LDOS and consider RG Eqs. \eqref{eq:ad:Z} and \eqref{eqmqRG}
in the intermediate range of parameters $|\gamma_0|\ll t_0\ll \sqrt{|\gamma_0|}$.
For temperatures $T\gg T_c$, the behavior of the disorder-averaged LDOS is fully determined by Eq. \eqref{eq:ad:Z}
expanded to the lowest order in interaction parameters.
After projection to the BCS line (\ref{BCSline}), we find from Eq. \eqref{eq:ad:Z}:
\begin{equation}
\frac{d \ln Z}{dy} = 4\gamma t .
\label{eq:Z:wsri}
\end{equation}
Solving this equation together with Eqs. \eqref{eq:1:e1} and \eqref{eq:1:e5}, we obtain
\begin{equation}
\frac{\langle \rho(E)\rangle}{\rho_0} = 1 + \gamma_0 \left ( \frac{t^2}{t_0^2}-1\right ) ,
\label{eq:rg:1a}
\end{equation}
where $t$ should be taken at the scale $\min\{L_T, L_E\}$.
Since we consider $T\gg T_c$, the resistivity $t(\min\{L_T, L_E\})\ll t_c$.
This allowed us to neglect the second term in the r.h.s of Eq. \eqref{eq:1:e5}.
As one can see from \eqref{eq:rg:1a}, there is a weak suppression of the averaged LDOS
for $t_0 \ll \sqrt{|\gamma_0|}$.
On the other hand, Eq. \eqref{eqmqRG} results in the following dependence of $m_q$ on $t$:
\begin{equation}
m_q = \bigl ( {t}/{t_0} \bigr )^{q(q-1)} .
\label{eq:1:e10}
\end{equation}
Here we omit the two-loop contribution, since within our approximation it is always
much smaller than the one-loop one.

Let us now consider temperatures close to the transition temperature,
\begin{equation}
T_c\text{Gi}\ll T-T_c \ll T_c.
\label{T-close-to-Tc}
\end{equation}
Here $\text{Gi}\sim t_c$ is the Ginzburg-Levanyuk number (see Ref. [\onlinecite{Konig2015}] for details).
The first condition here stems from the range of the applicability of the one-loop RG for the
coupling constants, $|\gamma_c|t<1$ [\onlinecite{Burmistrov2015b}], see discussion below
Eqs.~\eqref{eq:rg:final:t}-\eqref{eq:rg:final:Z}.

In addition to renormalization of the average LDOS in accordance to Eq. \eqref{eq:ad:Z},
there are fluctuation corrections due to Cooper channel interaction which stem from the real processes
(the processes with frequencies smaller than temperature).
Within the one-loop approximation, fluctuation corrections stem from the second
line of Eq. \eqref{rhores} with frequency and moment integration restricted to the
range $Dq^2, |\omega| \lesssim \max\{T_c,|E|\}$.
In order to take such fluctuation corrections accurately, one needs to renormalize the NLSM action
down to the energy scale $\max\{T, |E|\}$ and, then,
to evaluate the contribution in the second line of Eq. \eqref{rhores} with the renormalized parameters. \footnote{In fact, the renormalization of the resistivity $t$ continues below the energy scale $\max\{T, |E|\}$ and stops at the scale given by the dephasing rate $1/\tau_\phi$ which is generically smaller than $\max\{T, |E|\}$. In what follows, for simplicity, we will disregard the difference between these scales, which in 2D systems amounts to ignoring extra logarithmic factors of the order of $\ln(1/t)$ in corrections to $t$. }

This way, one finds
\begin{equation}
\frac{\langle \rho(E) \rangle}{\rho_0} = Z^{1/2}_{s}(T)
\Bigl [
1 - 8(1-\ln 2) t_c (T_c \tau_{GL})^2 \Bigr ] ,
\label{eq:1:e11a}
\end{equation}
for $|E| \ll \tau_{GL}^{-1}$,
\begin{equation}
\frac{\langle \rho(E) \rangle}{\rho_0} = Z^{1/2}_{s}(T)
\Bigl [ 1 + 2t_c \frac{T_c^2}{E^2} \ln (|E| \tau_{GL}) \Bigr ] ,
\label{eq:1:e11b}
\end{equation}
for $\tau_{GL}^{-1} \ll |E| \ll T_c$, and
\begin{eqnarray}
\frac{\langle \rho(E) \rangle}{\rho_0} &=&
Z^{1/2}_{s}(|E|)\left[1 + 2t(L_E) \frac{T_c^2}{E^2} \ln (T_c \tau_{GL})\right]
\nonumber
\\
&\simeq&
1 + 2t(L_E) \frac{T_c^2}{E^2} \ln (T_c \tau_{GL})  - \frac{3 t^2(L_E)}{2t_c} .
\label{eq:1:e11c}
\end{eqnarray}
for $T_c\ll |E| \ll 1/\tau$.
Here we have introduced the Ginzburg-Landau time as
\begin{equation}
\tau_{GL}^{-1} = 8 T_c |\gamma^{-1}(L_T)|/\pi.
\label{tauGL}
\end{equation}
For the standard BCS-type divergence of $\gamma_c$, e.g. as in Eq. \eqref{eq:1:e5},
$\tau_{GL}^{-1} \sim T-T_c$, as usual.
Expressions \eqref{eq:1:e11a}-\eqref{eq:1:e11c} were derived under assumption $T_c \tau_{GL} \gg 1$.

To find the renormalization factor $Z_s(T)$  (with the subscript `s' emphasizing that
the interaction is short-ranged), we should integrate the RG equations
\eqref{eq:rg:final:t}-\eqref{eq:rg:final:gc} and  \eqref{eq:ad:Z}. In the considered range of temperatures close to $T_c$, Eq.~\eqref{T-close-to-Tc}, this, strictly speaking,  cannot be done analytically, so that one should perform the integration numerically. In order to have an analytical approximation, we have used simplified RG equations \eqref{eq:1:e1}, \eqref{eq:1:e5}, and \eqref{eq:Z:wsri}, which yield
\begin{equation}
Z^{1/2}_s(T) = e^{3t(L_T)-3 t_0} \left [\frac{t_c-t(L_T)}{t_c-t_0}\right ]^{3 t_c}.
\label{Zs:def}
\end{equation}
 Since we consider
temperatures close to $T_c$, it is convenient to express $Z^{1/2}_s(T)$ via the Ginzburg-Landau time:
\begin{equation}
Z^{1/2}_s(T) = \left ( \frac{16}{3\pi e} \frac{T_c\tau_{GL}}{t_c}\right )^{-3 t_c}.
\label{eq:F1:est}
\end{equation}
We have checked that this analytical approximation is in reasonable agreement with the result of numerical integration of the exact RG equations.
The renormalization factor $Z^{1/2}_{s}(|E|)$ is obtained from Eq. \eqref{Zs:def} by substitution $L_E$ for $L_T$. The last term in the r.h.s. of Eq. \eqref{eq:1:e11c} is due to renormalization of $\langle \rho(E) \rangle$
from the energy scale $1/\tau$ down to $|E|$, in accordance with Eq. \eqref{eq:rg:1a}.

We mention that Eqs. \eqref{eq:1:e11a}-\eqref{eq:1:e11c} after the substitutions
$Z^{1/2}_s(T) \to 1$, $t(L_E)\to t_0$, and $t_c \to t_0$,
transform into the well-known results of a plain perturbation theory [\onlinecite{Abrahams1970,Maki1970,Castro1990}].
For temperatures close to $T_c$, renormalization effects due to interplay of disorder and interaction
yield the following two important novel features in comparison with perturbative results:
\begin{itemize}
\item[(i)] strong suppression of the disorder-averaged LDOS already at energies $|E| \sim T$ (see Eq. \eqref{eq:F1:est});
\item[(ii)] the Ginzburg-Levanyuk number is determined by the renormalized resistance $t_c$ rather than the bare one $t_0$, $\mathrm{Gi} \sim t_c$ (see Ref. [\onlinecite{Konig2015}] for details).
\end{itemize}

\begin{figure}[t]
	\centerline{\includegraphics[width=0.4\textwidth]{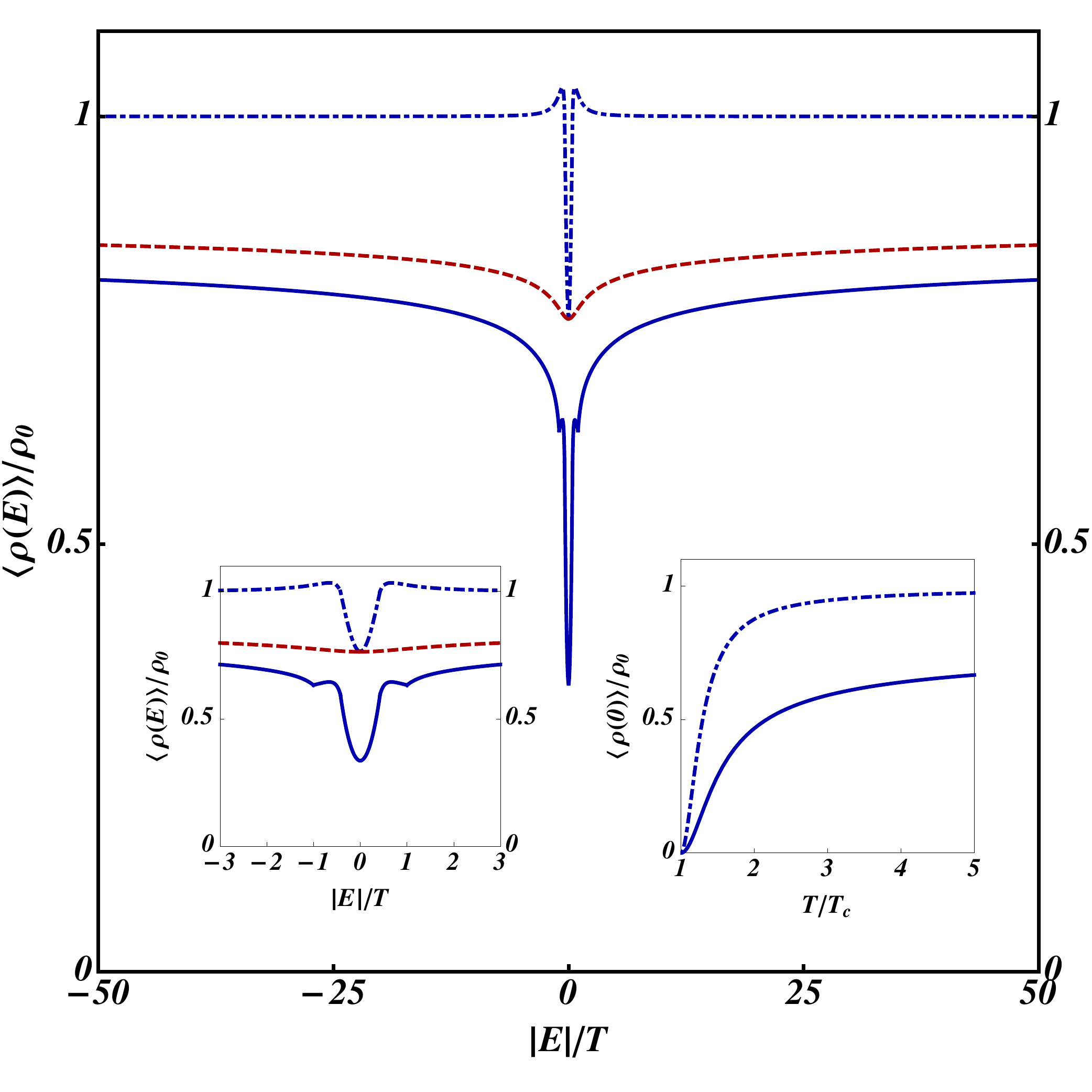}}
	\caption{(Color online) Sketch of the disorder-averaged LDOS $\langle \rho(E) \rangle$ in the case of short-range interaction for $t_c=0.05$ and $t_0 = 0.02$. The solid blue curve corresponds to the temperature $T=1.6 T_c$ ($T_c \tau_{GL} = 2.5$). The dashed red curve is plotted for $T=10T_c$. The dot-dashed blue curve is plotted for $T=1.6T_c$ according to expressions which ignore renormalization between the ultraviolet energy scale $1/\tau$ and $\max\{T,|E|\}$. The left inset: enlarged view of dependence of  $\langle \rho(E) \rangle$ on $|E|/T$. The right inset: dependence of  the disorder-averaged LDOS at $E=0$ on temperature.  See text and Eqs. \eqref{eq:rg:1a}, \eqref{eq:1:e11a} -- \eqref{eq:1:e11c} [\onlinecite{ftn1}].
}
\label{Fig:Fig1}
\end{figure}

The dependence of the disorder-averaged LDOS on energy is shown in Fig. \ref{Fig:Fig1}.
The  particularly strong energy dependence of the disorder-averaged LDOS appears at temperatures close to $T_c$.
It is worth emphasizing that the renormalization due to the interplay of disorder and interactions almost eliminates
coherence peaks, reducing them to a weak feature near $E \simeq \tau_{GL}^{-1}$. Up to these wiggles,
$\langle \rho(E) \rangle$ is a monotonously increasing function of energy
(compare the dotted and solid blue curves in Fig. \ref{Fig:Fig1}).
Further, the renormalization effects result in a pronounced reduction of the disorder-averaged LDOS in comparison with $\rho_0$ also at much higher temperatures, $T\gg T_c$  (see the dashed red curve in Fig. \ref{Fig:Fig1}).

The suppression of low-energy DOS due to renormalization factor (\ref{eq:F1:est}) becomes stronger when the disorder (i.e. $t_0$) increases, and the system approaches the SIT (where $t_c\sim 1$). This effect is further strengthened on the insulating side of SIT. The analysis is fully analogous to that performed below for the case of a system close to the Anderson transition, see Sec.~\ref{s5c} and the blue curve in Fig.~\ref{Fig:Fig5}. Physically, there is a close similarity between this effect and the development of the Coulomb zero-bias anomaly (ZBA) into soft Coulomb gap on the metal-insulator crossover (or transition) in normal systems [\onlinecite{Burmistrov2014}]. The difference is that in the present case the source of the renormalization-induced effects, the ZBA on the superconducting side and the soft gap on the insulating side, is the attractive interaction. 

We also note that the results \eqref{eq:1:e11a}-\eqref{eq:1:e11c} are obtained as the lowest order of the
expansion in the renormalized resistance.
It is sufficient to keep these lowest-order corrections only if we are not too close to the transition temperature, $T_c \tau_{GL} \ll 1/\sqrt{\mathrm{Gi}}\sim 1/\sqrt{t_c}$ [\onlinecite{Larkin2001},\onlinecite{Levchenko2010}], i.e. when the fluctuation correction
to the disorder-averaged LDOS is small. Importantly, the above condition justifying the perturbative treatment of the superconducting
fluctuation corrections to the density of states is stronger than the border $|\gamma_c|t\sim 1$ of validity
of the one-loop RG equations for the coupling constants. The situation is somewhat similar to what we encounter
in the case of Coulomb interaction, where the interaction corrections to the density of states (resulting from
effective gauge fluctuations) are stronger than the corrections to the conductivity, see Sec.~\ref{subsec:2DCoulomb} below.

\begin{figure}[t]
	\centerline{\includegraphics[width=0.4\textwidth]{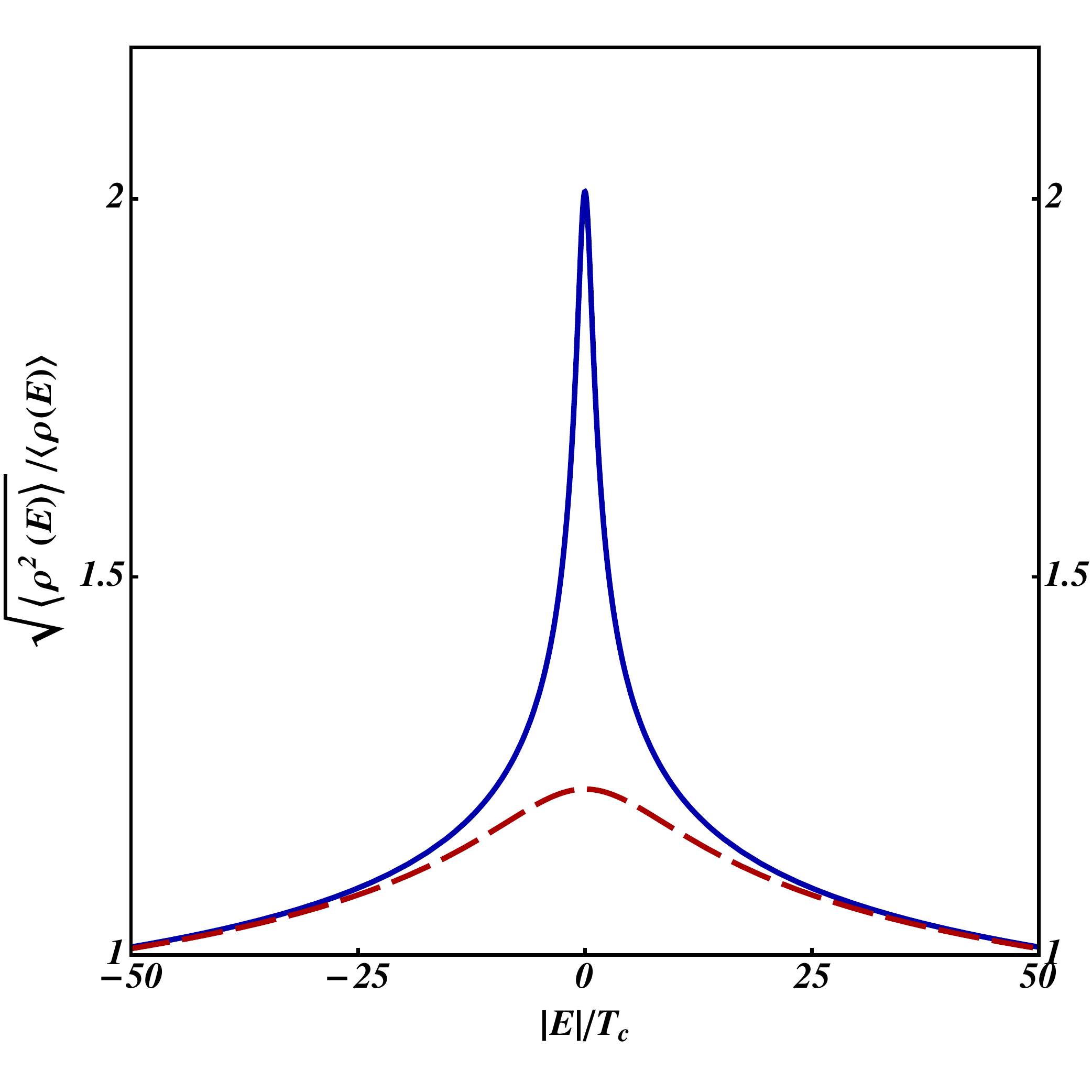}}
	\caption{(Color online) Ratio $\sqrt{\langle \rho^2(E) \rangle}/\langle \rho(E) \rangle$ characterizing the magnitude of LDOS fluctuations in the case of short-range interaction. The curves are obtained by the numerical solution of RG equations \eqref{eq:rg:final:t} - \eqref{eq:rg:final:gc} and \eqref{eqmqRG} with initial values $t_0=0.2$ and $\gamma_{c0} = \gamma_{t0}=-\gamma_{s0}=-0.2$. The solid blue curve corresponds to the temperature $T\simeq T_c$. The dashed red curve is plotted for $T=10T_c$.
}
\label{Fig:Fig1fl}
\end{figure}

Let us now analyze the fluctuations of the LDOS for short-ranged interactions.
Since the function $m_q$ (with $q\geqslant 2$) does not involve corrections from real processes,
within one-loop approximation its behavior at finite temperature and energy is fully determined
by RG result \eqref{eq:1:e10} with $t$ taken at the length scale $\min\{L_T, L_E\}$.
According to Eq. \eqref{eq:1:e10}, mesoscopic fluctuations of the LDOS are most pronounced at
temperatures close to the superconducting transition temperature, $T-T_c\ll T_c$.
Using Eqs. \eqref{eq:1:e1} and \eqref{eq:1:e10} we obtain
\begin{equation}
\frac{\langle \rho^q(E) \rangle}{\langle \rho(E) \rangle^q} =  \left (\frac{t_c}{t_0}\right )^{q(q-1)}
\label{eq:1:e12a}
\end{equation}
for $|E| \ll T_c$ and
\begin{equation}
\frac{\langle \rho^q(E) \rangle}{\langle \rho(E) \rangle^q} =  \left (\frac{t_c}{t_0}\right )^{q(q-1)}
 \left ( 1+ \frac{t_c}{2}\ln \frac{E}{T_c} \right )^{q(1-q)}
\label{eq:1:e12b}
\end{equation}
for  $T_c \ll |E|$. This result implies that at energies $|E| < T_c$ fluctuations of the LDOS are large and non Gaussian.
For $|E|>T_c$ their amplitude decreases with increasing energy.
The LDOS fluctuations are particularly strong for weak initial couplings satisfying $|\gamma_0|\ll t_0\ll \sqrt{|\gamma_0|}$ (which is the range of parameters where the enhancement of $T_c$ takes place), since in this case $t_c \gg t_0$. The fluctuations remains strong, for $T$ close to $T_c$, also for intermediate initial couplings, as found from numerical solution of the full RG equations. To illustrate this, we show  in Fig. \ref{Fig:Fig1fl} the ratio $\sqrt{\langle \rho^2(E) \rangle}/\langle \rho(E) \rangle$ obtained by numerical integration of Eqs.~\eqref{eq:rg:final:t} - \eqref{eq:rg:final:gc} and Eq.~\eqref{eqmqRG} for $q=2$ (neglecting the two-loop contribution in Eq. \eqref{eqmqRG}).

Finally, we note in passing that there is a fluctuation correction to $m_q$ in the two-loop approximation.
The two-loop correction to $[P_2^{\alpha_1\alpha_2}]^{RA}$ (see Eq. \eqref{eq2loopK2_P2+-1})
can be cast as renormalization of the one-loop mesoscopic diffuson and cooperon (see Appendix \ref{app:sec:2}).
Therefore, it does not produce fluctuation corrections to $m_q$.
The two-loop renormalization of $[P_2^{\alpha_1\alpha_2}]^{RR}$ contains the
contribution due to real processes in the Cooper channel (see Eq. \eqref{eq2loopK2_P2++11}).
However, as one can check, it leads to a correction which is by factor $t_c \ln T_c \tau_{GL} \ll 1$
smaller than the fluctuation corrections to the disorder-averaged LDOS.

\subsection{Coulomb interaction in 2D}
\label{subsec:2DCoulomb}

Let us now consider the case of Coulomb interaction combined
with weak attraction in the Cooper channel.
For simplicity, we assume that the spin-rotational symmetry
is fully broken such that the triplet channel is absent, $\mathcal{N}=0$.
Then the full set of RG Eqs. \eqref{eq:rg:final:t} - \eqref{eq:rg:final:gc}
can be reduced to the following system of two equations [\onlinecite{Castellani1984b},\onlinecite{Ma1986}]:
\begin{equation}
\frac{dt}{dy} = \frac{t^2}{2},  \qquad
\frac{d\gamma_c}{dy} = \frac{t}{2} - 2\gamma_c^2 .
\label{eq:RG:SO:C}
\end{equation}
Here we neglect terms, which are powers of $\gamma_c$, as compared to unity. Equations \eqref{eq:RG:SO:C} predicts that in the weak coupling region, $t_0, |\gamma_{c,0}| \ll 1$, superconductivity exists for $t_0<4\gamma_{c,0}^2$ . The Cooper channel attraction becomes of the order of unity at the length scale [\onlinecite{Finkelstein1987}]
\begin{equation}
y_{c} = \frac{1}{2\sqrt{t_0}} \ln \frac{2|\gamma_{c,0}|+\sqrt{t_0}}{2|\gamma_{c,0}|-\sqrt{t_0}}  \gg 1.
\end{equation}

We first consider the region of the phase diagram away from the separatrix $t_0 = 4\gamma_{c,0}^2$,
when the following inequality holds: $y_c \ll 1/t_0$.
The resistance $t_c = t(y_c)$ at this scale is not very different from $t_0$:
\begin{equation}
t_c = t_0 \left  ( 1 + \frac{\sqrt{t_0}}{4} \ln \frac{2|\gamma_{c,0}|+\sqrt{t_0}}{2|\gamma_{c,0}|-\sqrt{t_0}}\right )  \sim t_0 \ll 1 .
\label{eq:t:c:1}
\end{equation}
Therefore, one can expect that beyond the length scale $y_c$ the superconducting instability
develops under RG flow. The transition temperature to the superconducting phase can be
estimated as $T_{c} \sim \tau^{-1} \exp(-2 y_c)$. Due to the combined effect of disorder
and Coulomb interaction, the transition temperature $T_{c}$ decreases with growing
$t_0$ from $T_c^{\rm BCS}$ at $t_0=0$ to $0$ at $t_0 = 4\gamma_{c,0}^2$ [\onlinecite{Finkelstein1987}].

In the considered case of Coulomb interaction, the disorder-averaged LDOS is dominated
by the first term (which is formally infinite for $\gamma_s=-1$) in the r.h.s of Eq. \eqref{eq:ad:Z}.
In fact, this means that the disorder-averaged LDOS is strongly suppressed by gauge-type fluctuations 
[\onlinecite{Altshuler1980},\onlinecite{Finkelstein1983}].
Taking into account corrections due to interaction in the Cooper
channel [\onlinecite{Abrahams1970},\onlinecite{Maki1970},\onlinecite{Castro1990}]
which become important only at temperatures close to the superconducting transition
temperature, $T-T_c\ll T_c$, we find
\begin{equation}
\frac{\langle \rho(E) \rangle}{\rho_0} =
Z^{1/2}_c(T_c) \Bigl [ 1 - 8(1-\ln 2) t_c (T_c \tau_{GL})^2 \Bigr ],
\label{eq:t:c:3a}
\end{equation}
for $|E| \ll \tau_{GL}^{-1}$,
\begin{equation}
\frac{\langle \rho(E) \rangle}{\rho_0} =
Z^{1/2}_c(T_c) \Bigl [1 +  2\frac{t_c T_c^2}{E^2} \ln (|E| \tau_{GL})\Bigr ] ,
\label{eq:t:c:3b}
\end{equation}
for $\tau_{GL}^{-1} \ll |E| \ll T_c$, and
\begin{equation}
\frac{\langle \rho(E) \rangle}{\rho_0} =
Z^{1/2}_c(|E|) \Bigl [1 +  2\frac{t(L_E) T_c^2}{E^2} \ln (T_c \tau_{GL})\Bigr ],
\label{eq:t:c:3c}
\end{equation}
for $T_c\ll |E|$. Here the function $Z^{1/2}_c(|E|)$ describes the zero-bias anomaly for Coulomb interaction (hence the subscript `c')
and is given by [\onlinecite{Altshuler1980},\onlinecite{Finkelstein1983},\onlinecite{Nazarov1989},\onlinecite{LevitovShytov1997}]
\begin{equation}
Z^{1/2}_c(|E|) = \exp \left  [ - \frac{t_0}{4} \ln \bigl (E \tau \bigr )  \ln
\frac{E \tau}{\varkappa^2 l^2}\right ] ,
\label{Zc:def}
\end{equation}
where $\varkappa = (2\pi e^2/\varepsilon) dn/d\mu$ stands for the inverse static
screening length in a 2D electron system. The renormalization factor $Z^{1/2}_c(T_c)$ is obtained by substitution of $T_c$ for $|E|$ in Eq. \eqref{Zc:def}. It can be estimated as
\begin{align}
Z^{1/2}_c(T_c) = & \exp \Biggl  [ - \frac{1}{4} \ln \left ( \frac{2|\gamma_{c,0}|-\sqrt{t_0}}{2|\gamma_{c,0}|+\sqrt{t_0}} \right )
\notag
\\
& \times \ln \left (
\frac{2|\gamma_{c,0}|-\sqrt{t_0}}{2|\gamma_{c,0}|+\sqrt{t_0}}(\varkappa l)^{-2\sqrt{t_0}}\right) \Biggr ] \ll 1.
\label{Zc:def1}
\end{align}

At temperatures $T\gg T_c$ the disorder-averaged LDOS is fully determined by the zero-bias anomaly,
\begin{equation}
{\langle \rho(E) \rangle}/{\rho_0} = Z^{1/2}_c(\max\{|E|,T\}), \qquad T\gg T_c.
\label{ZBAc}
\end{equation}

We note that Eqs. \eqref{eq:t:c:3a}-\eqref{eq:t:c:3c} after the following substitutions: $
Z^{1/2}_c \to 1$, $t(L_E)\to t_0$, and $t_c \to t_0$, transform into the well-known results of
naive perturbation theory [\onlinecite{Abrahams1970,Maki1970,Castro1990}]. It is also worth mentioning that the
renormalization of the disorder-averaged LDOS due to attraction in the Cooper channel [the
term proportional to $\gamma_c$ in Eq. \eqref{eq:ad:Z}] is not important. As one can check, the Cooper channel term yields the contribution to $\ln Z^{1/2}_c(T_c)$ which is the factor $\sqrt{t_0} y_c \ll 1$ smaller than the zero bias anomaly contribution \eqref{Zc:def1}.

\begin{figure}[t]
	\centerline{\includegraphics[width=0.4\textwidth]{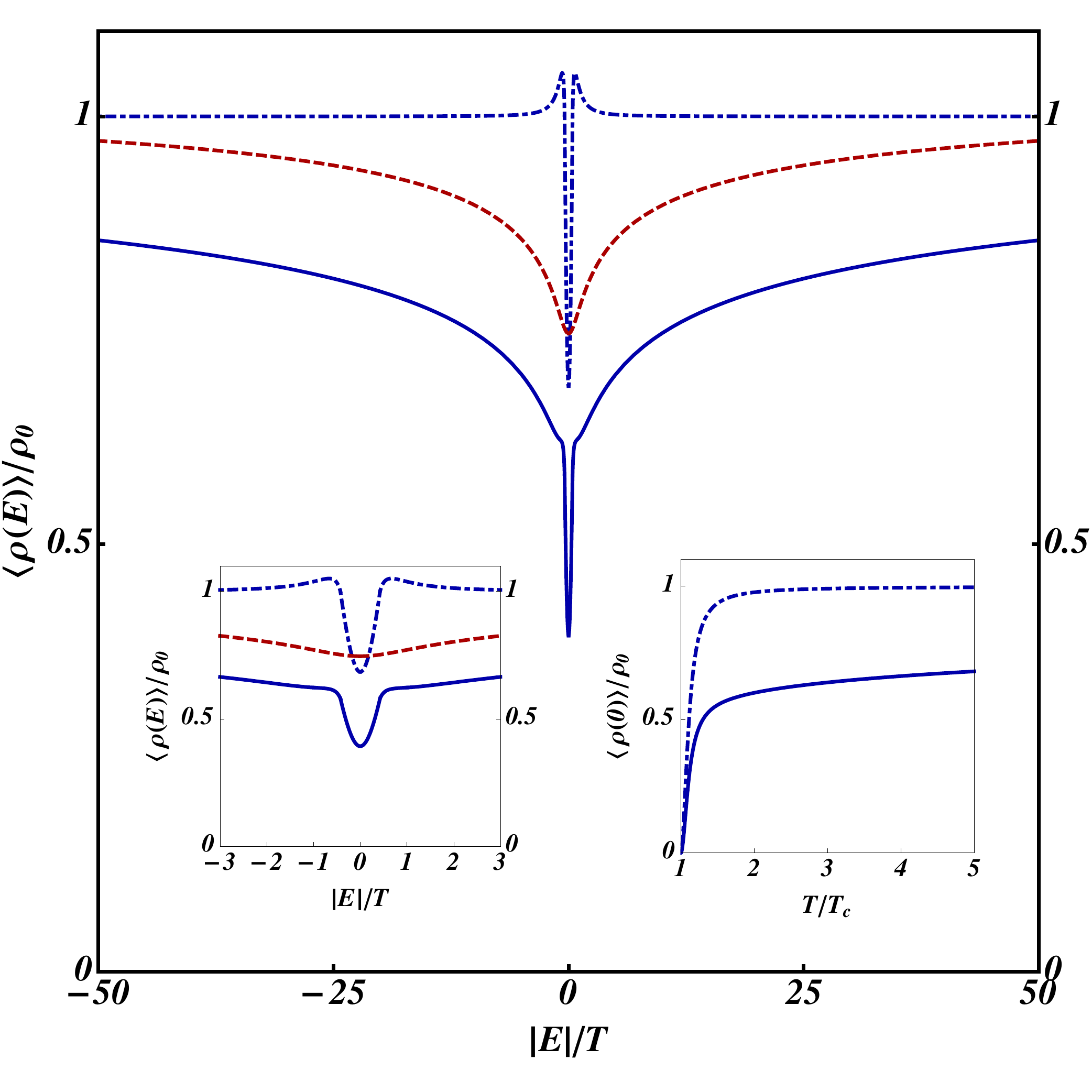}}
	\caption{(Color online) Sketch of the disorder-averaged LDOS $\langle \rho(E) \rangle$ in the case of Coulomb interaction for $t_0=0.03$. The solid blue curve corresponds to the temperature $T=1.2 T_c$ ($T_c \tau_{GL} = 2.5$). The dashed red curve is plotted for $T=10T_c$. The dot-dashed blue curve is plotted for $T=1.2T_c$ according to expressions which ignore renormalization between the ultraviolet energy scale $1/\tau$ and $\max\{T,|E|\}$. The left inset: enlarged view of dependence of  $\langle \rho(E) \rangle$ on $|E|/T$. The right inset: dependence of  the disorder-averaged LDOS at $E=0$ on temperature. We choose $\varkappa l =0.2$. See text and Eqs. \eqref{eq:t:c:3a}-\eqref{eq:t:c:3c} [\onlinecite{ftn1}].}
	\label{Fig:Fig2}
\end{figure}

The dependence of the disorder-averaged LDOS on energy is shown in Fig.~\ref{Fig:Fig2}.
The pronounced energy dependence of the disorder-averaged LDOS appears at temperatures close to $T_c$.
The overall behavior of $\langle \rho(E) \rangle$  with energy is similar to the case of short-ranged
interaction (see Fig. \ref{Fig:Fig2}). At the same time, we emphasize the important difference between the two cases.
For short-ranged interaction, the suppression of $\langle \rho(E) \rangle$ both at energies $|E| \gtrsim T_c$ and
at $|E| \lesssim \tau_{GL}^{-1}$ is controlled by the attractive interaction.
Contrary to this, in the case of Coulomb interaction, the suppression of LDOS at energies $|E| \gtrsim T_c$ is
dominated by the zero-bias anomaly factor $Z^{1/2}_c(E)$, whereas at $|E| \lesssim \tau_{GL}^{-1}$
the suppression of $\langle \rho(E) \rangle$ is due to Cooper channel attraction.
It is worth mentioning that, parametrically, the Coulomb renormalization factor \eqref{Zc:def} may lead to a much stronger suppression of LDOS than  the attractive-interaction renormalization factor \eqref{eq:F1:est}. On the other hand, for realistic parameters this difference is usually not so dramatic, cf. Figs. \ref{Fig:Fig1} and \ref{Fig:Fig2}.

The two-loop result \eqref{eqmqRG} for the anomalous dimension of the $q$-th moment of LDOS
for the case of Coulomb interaction takes the following form:
\begin{equation}
\frac{d \ln m_q}{d y} = \frac{q(q-1)}{8} \Bigl  [2 t +  \bigl (2-\frac{\pi^2}{6} - 2\gamma_c\bigr ) t^2 \Bigr ]  .
\label{eqmqRG:SO}
\end{equation}
In the considered case, one can neglect the two-loop contribution.
Using Eq. \eqref{eq:RG:SO:C}, we find the following one-loop RG result from Eq. \eqref{eqmqRG:SO}
\begin{equation}
m_q = (t/t_0)^{q(q-1)/2} .
\label{mq:CI}
\end{equation}
Hence for the disorder-averaged moments of LDOS
at temperatures close to the superconducting temperature, $T-T_c\ll T_c$:
\begin{equation}
\frac{\langle \rho^q(E) \rangle}{\langle \rho(E) \rangle^q} =  \left (\frac{t_c}{t_0}\right )^{\frac{q(q-1)}{2}}
\label{eq:t:c:2a}
\end{equation}
for $|E| \ll T$ and
\begin{equation}
\frac{\langle \rho^q(E) \rangle}{\langle \rho(E) \rangle^q} =  \left (\frac{t_c}{t_0}\right )^{\frac{q(q-1)}{2}}
 \left ( 1 + \frac{t_c}{4}\ln \frac{E}{T_Ó} \right )^{\frac{q(1-q)}{2}}
 \label{eq:t:c:2b}
\end{equation}
for $T \ll |E|$. 

\begin{figure}[t]
	\centerline{\includegraphics[width=0.4\textwidth]{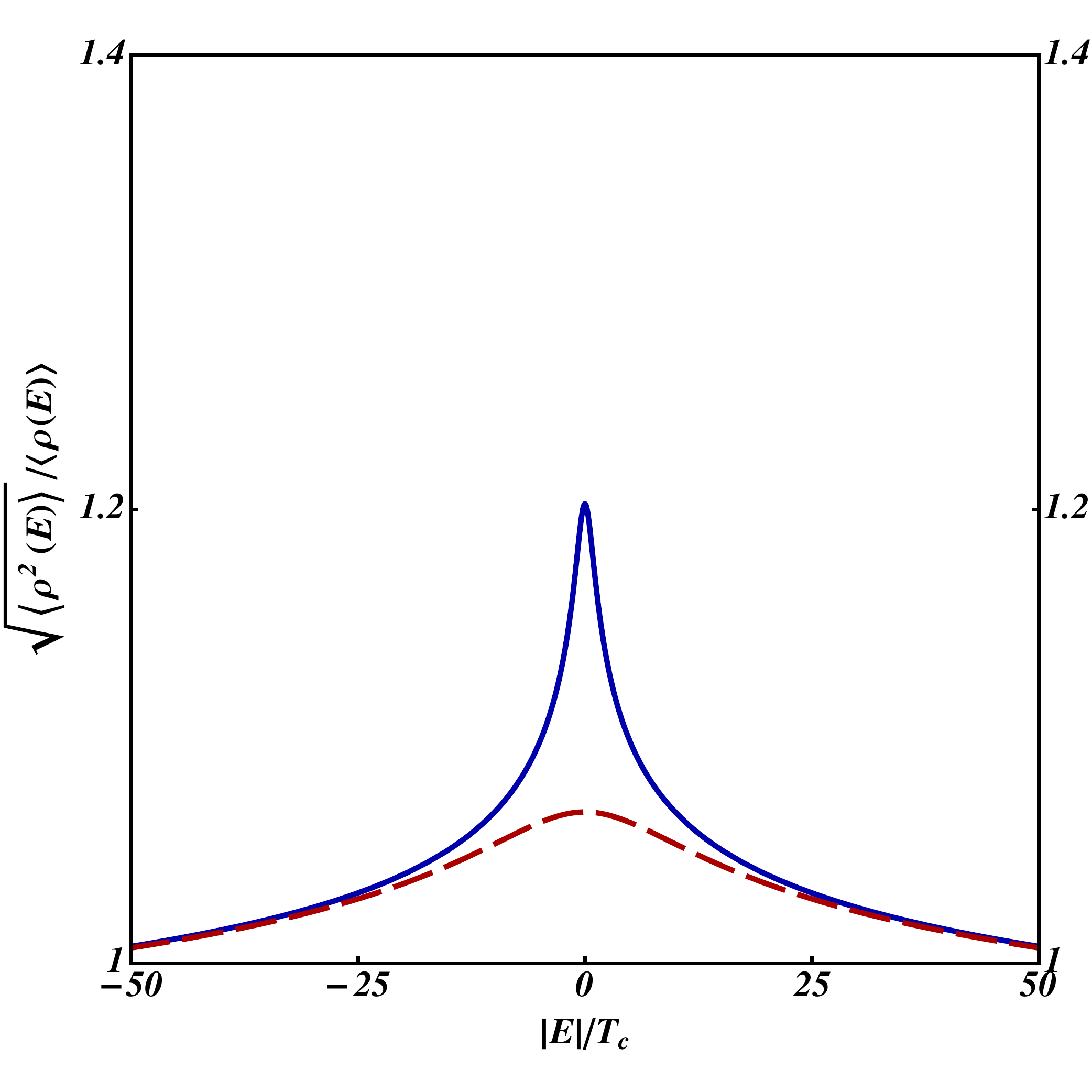}}
	\caption{(Color online) 
Ratio $\sqrt{\langle \rho^2(E) \rangle}/\langle \rho(E) \rangle$ characterizing the magnitude of LDOS fluctuations in the case of Coulomb interaction. The curves are obtained by the numerical solution of RG equations \eqref{eq:rg:final:t} - \eqref{eq:rg:final:gc} (with $\mathcal{N}=0$) and Eq.\eqref{eqmqRG} with initial values $t_0=0.25$ and $\gamma_{c0} =-0.35$. 
The solid blue curve corresponds to the temperature $T\simeq T_c$. The dashed red curve is plotted for $T=10T_c$.	
}
\label{Fig:Fig2fl}
\end{figure}

When the initial couplings are weak, then, according to Eq. \eqref{eq:t:c:1}, $t_c$ is typically close to $t_0$, exceeding it only slightly.
Therefore, the mesoscopic fluctuations of LDOS in a problem with Coulomb interaction are in general weak for weak bare couplings. The only exception is a vicinity of the SIT separatrix, $t_0=4\gamma_{c,0}^2$, such that 
$$
\ln\frac{2\sqrt{t_0}}{2|\gamma_{c,0}|-\sqrt{t_0}} \gtrsim 1/\sqrt{t_0}.
$$
In this regime $t_c \gg t_0$, so that the LDOS fluctuations  \eqref{eq:t:c:2b} become parametrically strong.  This should be contrasted with the case of short-range interaction, for which there is a parametrically broad regime, $|\gamma_0|\ll t_0\ll \sqrt{|\gamma_0|}$, of very strong fluctuations, see Sec.~\ref{s4b}.
 For intermediate values of bare couplings, the LDOS fluctuations can be quantified, in full analogy with the short-range-interaction case, by a numerical solution of the RG equations. In Fig.~\ref{Fig:Fig2fl} we display typical results obtained by numerical integration of Eqs. \eqref{eq:rg:final:t} - \eqref{eq:rg:final:gc} (with $\mathcal{N}=0$) and Eq. \eqref{eqmqRG-SO} (with $q=2$ and neglecting the two-loop contribution). A comparison of Figs.  \ref{Fig:Fig1fl} and \ref{Fig:Fig2fl} demonstrates that the LDOS fluctuations in the Coulomb case are weaker than in the short-range case with similar parameters.

At $T\gg T_c$ the $q$-th moment of LDOS is given by Eq. \eqref{mq:CI} with $t$ taken at the length scale $\min\{L_T,L_E\}$.

Recently, mesoscopic fluctuations of the cooperon have been analyzed
[\onlinecite{SkvortsovFeigelman2005}]. This study has revealed that the typical scale
of fluctuations of the transition temperature to the superconducting state
can be estimated as $\delta T_c/T_c \sim t_0^2\gamma_{c,0}^2/(4\gamma_{c,0}^2-t_0)$.
Near the separatrix, $4\gamma_{c,0}^2-t_0\ll t_0^3$,
the fluctuations of the transition temperature becomes large, $\delta T_c/T_c \gg 1$.
We mention that the mesoscopic fluctuations of the LDOS remain small
at $t_0 \exp(-1/\sqrt{t_0})\lesssim 4\gamma_{c,0}^2-t_0\ll t_0^3$.

\section{System at or near a noninteracting Anderson transition}
\label{s5}

Let us now consider the case of weak short-range interactions, assuming
that in the absence of interactions the system is near the Anderson transition.
The physical examples include a 2D electron system with broken spin-rotational
symmetry (class AII) or 3D electron systems with preserved or broken spin-rotational
symmetry (classes AI and AII, respectively).
For the sake of concreteness, we assume that the spin rotational symmetry is not broken.

Since we assume that we start from weak interactions,
the term describing the usual BCS instability is not important initially.
Then, similar to Eqs \eqref{eq:1:e2}, during the first part of the RG evolution
the coupling constants $\gamma_s$  and $\gamma_t$ adjust themselves
to $\gamma_c$ according to $\gamma_s = - \gamma_t = - \gamma_c \equiv -\gamma$ (the BCS line).
Then the superconducting instability can be described by single equation for $\gamma$:
\begin{equation}
\label{eq:AT:e4.1}
\frac{d\gamma}{ dy} = -\Delta_2 \gamma - a\gamma^2 .
\end{equation}
Here $\Delta_2<0$ stands for the multifractal exponent of the inverse
participation ratio (bilinear in $Q$ operator $K_2$) at the noninteracting
fixed point describing the Anderson transition.
For the case of 3D Anderson transition in class AI numerical simulations yield the
estimate $\Delta_2=-1.7 \pm 0.05$ [\onlinecite{Mildenberger2002}].
The constant $a$ is a universal number which is determined by the properties of composite operators at the noninteracting fixed point (cf. Ref. [\onlinecite{Burmistrov2011}]). As follows from Eqs. \eqref{eq:rg:final:gs} - \eqref{eq:rg:final:gc}, for $t\ll 1$ (i.e., at the critical point in $2+\epsilon$ dimensions), the coefficient $a$ is positive, $a=2/3-t>0$.

We note that Eq. \eqref{eq:AT:e4.1} is written under the assumption that the attraction is weak, $|\gamma|\ll 1$.
As one can see from Eq. \eqref{eq:rg:final:gc}, disorder generates higher powers of $\gamma$ on the right-hand
side of Eq. \eqref{eq:AT:e4.1}. In our analysis below, we assume that Eq. \eqref{eq:AT:e4.1} describes
the renormalization of $\gamma$ both in the limit of weak and large attraction (see Ref. [\onlinecite{Konig2015}] for details).
In other words, we assume that even in the presence of disorder the superconducting instability occurs via the BCS scenario.
This implies $a>0$ in Eq. \eqref{eq:AT:e4.1}. We emphasize that, within this assumptions,
the analysis below is not sensitive to the details of the RG equation for intermediate values of $\gamma$.

Interactions affect the renormalization of the resistivity near the Anderson transition point.
The corresponding RG equation for $\gamma\ll 1$ can be written as follows:
\begin{equation}
\frac{d t}{dy} =\frac{1}{\nu}(t-t_\ast) + \eta \gamma . \label{eq:AT:e4.2a}
\end{equation}
Here, $t_\ast \sim 1$ is the critical value of the resistivity in the noninteracting case.
The second term with $\eta$ generalizes the Altshuler-Aronov conductivity correction.
In Eq. \eqref{eq:AT:e4.2a} we take in account that the presence of interactions drives the system
away from the non-interacting critical point.
In analogy with the coefficient $a$ in Eq.~\eqref{eq:AT:e4.1}, the constant  $\eta$ is a universal characteristics of the non-interacting fixed point.
In general, $\eta$ can be a positive or negative number of the order unity.
The non-interacting correlation length exponent is positive, $\nu>0$.
For the 3D Anderson transition in the presence of spin rotational symmetry numerical simulations yield the value $\nu= 1.57\pm 0.02$  [\onlinecite{Slevin1999},\onlinecite{Slevin2014}].

We now proceed in the same two-step manner, as before. For initially weak interaction,
neglecting the term of the second order in $\gamma$,
Eqs.~\eqref{eq:AT:e4.1} and \eqref{eq:AT:e4.2a} can be solved.
For $|\Delta_2|\nu \neq 1$ we find
\begin{equation}
t = \tilde t + \frac{\eta \nu}{|\Delta_2|\nu-1} \gamma,
\end{equation}
where
\begin{equation}
\qquad
\tilde t =t_\ast + \Bigl ( \tilde t_0-t_\ast\Bigr )e^{y/\nu} , \quad
\gamma=\gamma_0 e^{|\Delta_2| y},
\end{equation}
and
\begin{equation}
\tilde t_0 = t_0 -  \frac{\eta \nu\gamma_0}{|\Delta_2|\nu-1} .
\end{equation}
Therefore, in the presence of attractive interaction, the proper scaling
variable is $\tilde t$ rather than $t$.
However, since we assume that $|\gamma_0| \ll 1$ whereas $t_0 \sim t_\ast \sim 1$
the difference between $t_\ast$ and $\tilde{t}_c = t_\ast - \eta \nu\gamma_0/(|\Delta_2|\nu-1)$ is small.

In the special case $|\Delta_2|\nu=1$, we find
\begin{equation}
\tilde t = t - \eta \gamma_0 \gamma y, \quad \tilde t_0 = t_0.
\end{equation}
In what follows we shall omit `tilde' sign and neglect the difference between $\tilde{t}_\ast$ and $t_\ast$.

At finite temperature, the RG flow given by Eq. \eqref{eq:AT:e4.1} terminates at the length scale $L_T = l (T \tau)^{-1/d}$.
Solving Eq. \eqref{eq:AT:e4.1}, we find the following estimate for the temperature of transition to the superconducting state:
\begin{equation}
T_c^* = \tau^{-1} e^{-d y_c} = \frac{1}{\tau}\left ( \frac{a|\gamma_0|}{|\Delta_2|}\right )^{d/|\Delta_2|}
\label{eq:tcc}
\end{equation}

For $t<t_\ast$ the resistance decreases under RG flow.
At the scale $\xi = l |t_0/t_\ast-1|^{-\nu}$ the resistance vanishes,
whereas the Cooper-channel interaction becomes
\begin{equation}
\label{eq:AT:e4.5}
\gamma(\xi) = \gamma_{0} (\xi/l)^{|\Delta_2|} \,.
\end{equation}
After this (in view of $t=0$), the further renormalization of $\gamma$ is controlled by
the standard disorder-free BCS mechanism.
Then the transition temperature can be estimated as
\begin{equation}
T_c(\xi) = \delta_\xi e^{-1/|\gamma(\xi)|}
= \delta_\xi \exp \left [ - \frac{a}{|\Delta_2|} \left (\frac{T_c^*}{\delta_\xi}\right )^{\Delta_2/d} \right ] ,
\label{eq:AT:e4.7}
\end{equation}
where $\delta_\xi = \tau^{-1}(\xi/l)^{-d}$ stands for the typical level spacing
in the volume of linear size $\xi$.
This expression for $T_c(\xi)$ interpolates between $T_c{(\xi)} \sim T_c^*$ at $\delta_\xi
\sim T_c^*$  and $T_c{(\xi)} \sim T_c^{\rm BCS}$ at $\delta_\xi \sim 1/\tau$.

\begin{figure}[t]
\begin{tabular}{c}
\includegraphics[width=0.475\textwidth]{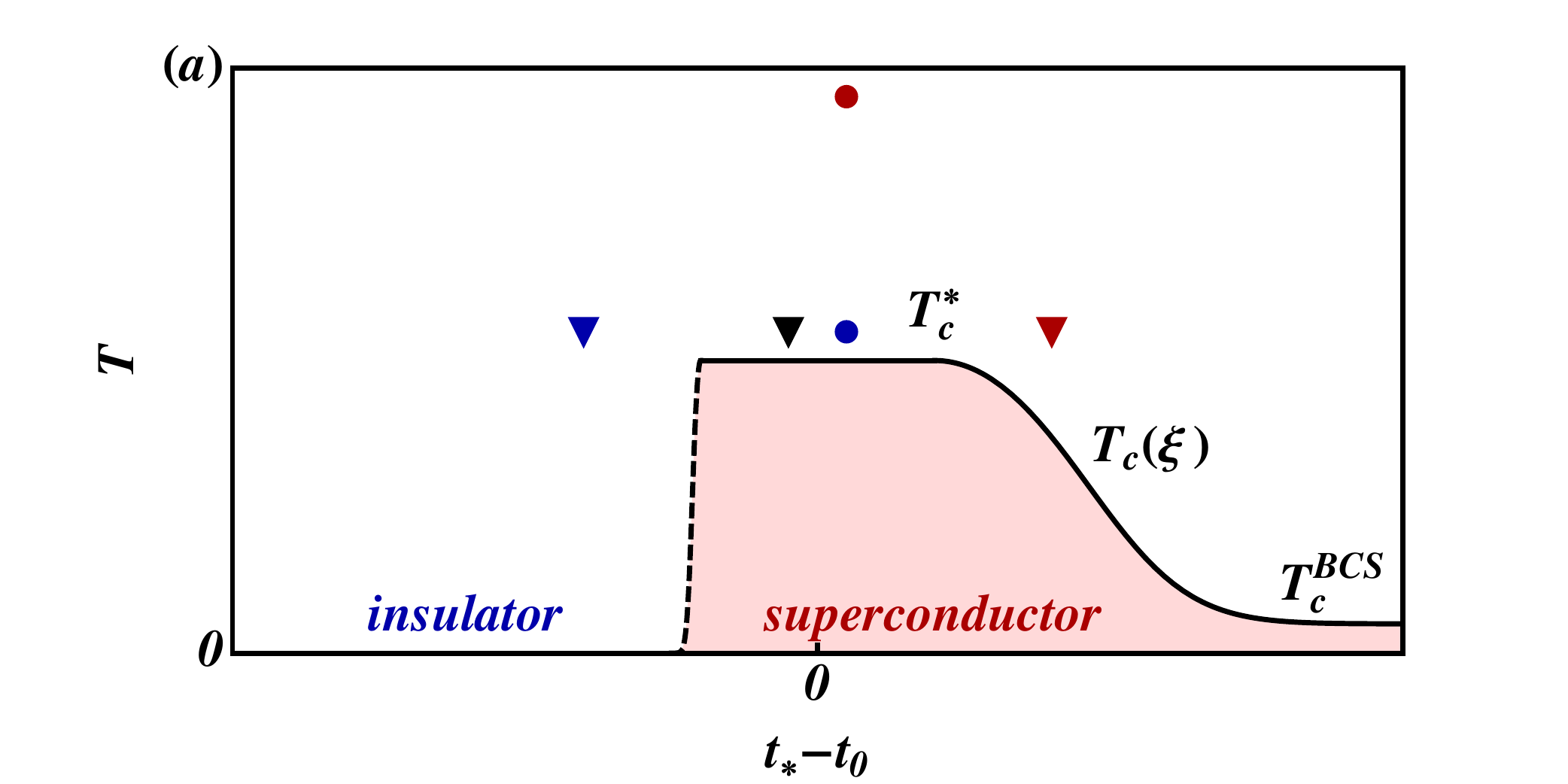} \\
\includegraphics[width=0.475\textwidth]{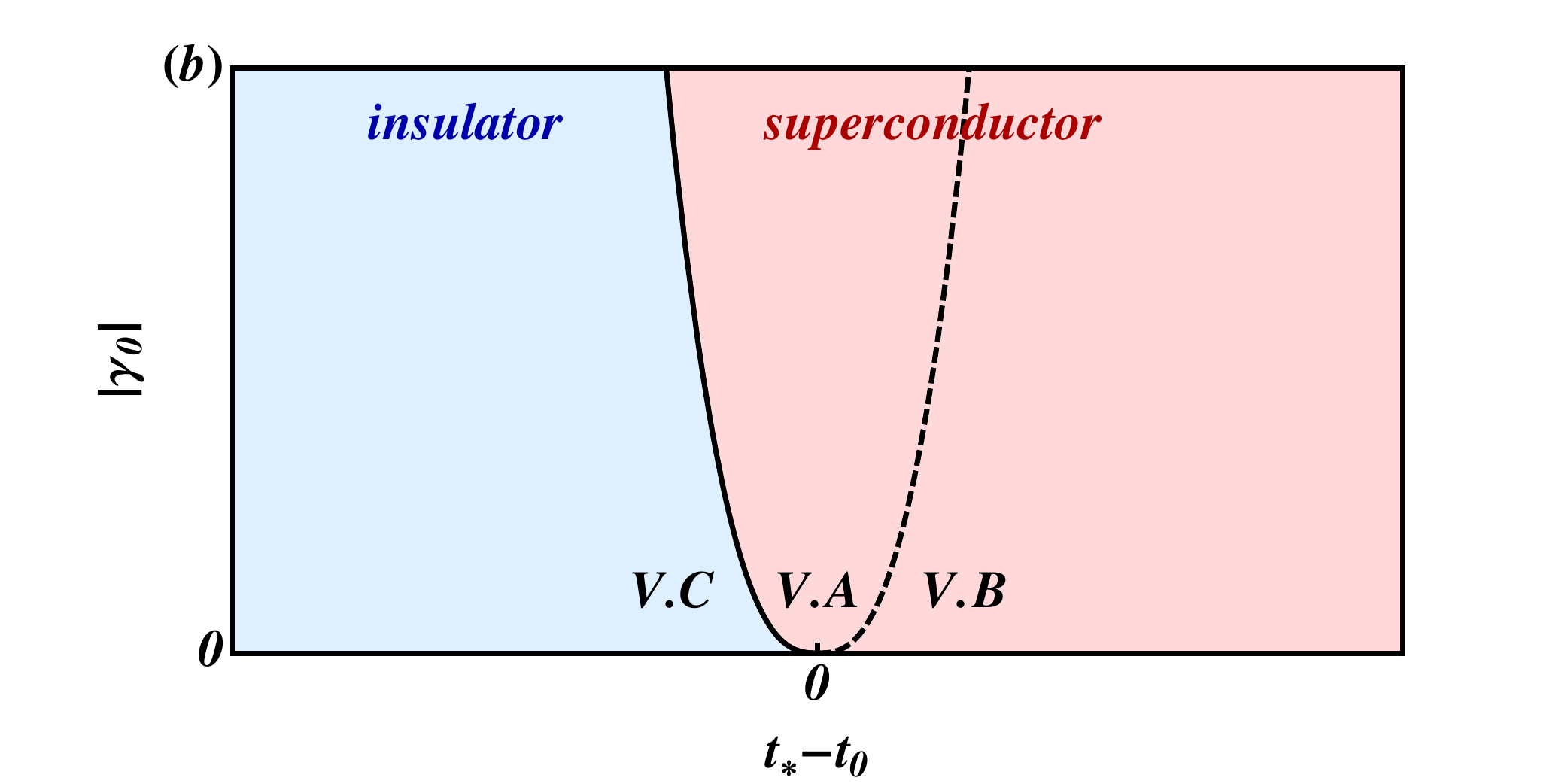}
\end{tabular}
\caption{(Color online) (a) Sketch of the superconducting transition
temperature as a function of distance from the critical point, $t_0-t_\ast$, at fixed value of bare attraction, $\gamma_0$,  (see Eqs. \eqref{eq:tcc} and \eqref{eq:AT:e4.7}). The red and blue dots correspond to the temperatures at which the energy dependences of LDOS in Fig. \protect\ref{Fig:Fig4} are shown.
 The blue, black and red triangles corresponds to curves $\langle \rho(E)\rangle$ shown  in Fig. \protect\ref{Fig:Fig5}. (b) Schematic phase diagram in the interaction-disorder plane near the critical point. Solid black curve denotes the superconductor-insulator transition. The symbols `V.A', `V.B' and `V.C' mark regions of the phase diagram in which behavior of LDOS is discussed in corresponding sections.
}
\label{Fig:Fig3}
\end{figure}

For $t>t_\ast$, in the absence of attraction the system would be in the insulating phase.
In the presence of interaction in the Cooper channel, the RG flow proceeds in accordance
with Eqs.~\eqref{eq:AT:e4.1} and \eqref{eq:AT:e4.2a} up to the localization length $\xi$.
There are two possibilities. If $L_*=l\exp(y_c) < \xi$, the attraction becomes of the
order of unity in the critical region of noninteracting Anderson transition.
Then one expects that the superconducting phase establishes below temperature $T_c^*$.
In the opposite case, $L_*>\xi$, the localization takes place first, and superconductivity
is not developed. Thus the relation
\begin{equation}
\label{eq:AT:e4.4}
\delta_\xi \sim T_c^* \, \quad \Leftrightarrow \quad |\gamma_0| \sim (t_\ast-t_0)^{\nu|\Delta_2|}
\end{equation}
is the condition of the superconductor-insulator transition (at zero temperature) (see Fig.~\ref{Fig:Fig3}).
It is worth noting that Ref.~[\onlinecite{FeigelmanCuevas2010}] argues that superconducting
state with $T_c \ll T_c^*$ persists further in the localized regime  ($\delta_\xi > T_c^*$)
due to Mott-type rare configurations. Our RG approach (at least, in its present form)
is not sufficient to explore this possibility.

\subsection{Exactly at criticality}
\label{s5a}

On the BCS line \eqref{BCSline},  the renormalization of the local density of states
near the Anderson transition fixed point can be described by the following equation:
\begin{equation}
\frac{d\ln \rho}{d y} = b \gamma,
 \label{eq:AT:LDOS1}
 \end{equation}
 where the coefficient $b$ is determined by the
 scaling properties of the Finkelstein term
 at the noninteracting fixed point.
As follows from Eq. \eqref{eq:Z:wsri}, at $t\ll 1$, the coefficient $b$ is positive, $b =2 t >0$.
Below we assume that it is positive in general.
We emphasize that there is no renormalization of the local density of states
in the absence of interaction. Renormalization group Eq. \eqref{eq:AT:LDOS1} is stopped at
the length scale $\mathcal{L} = \min\{L_T, L_E\}$ where $L_E = l (|E|\tau)^{-1/d}$.

\begin{figure}[t]
	\centerline{\includegraphics[width=0.4\textwidth]{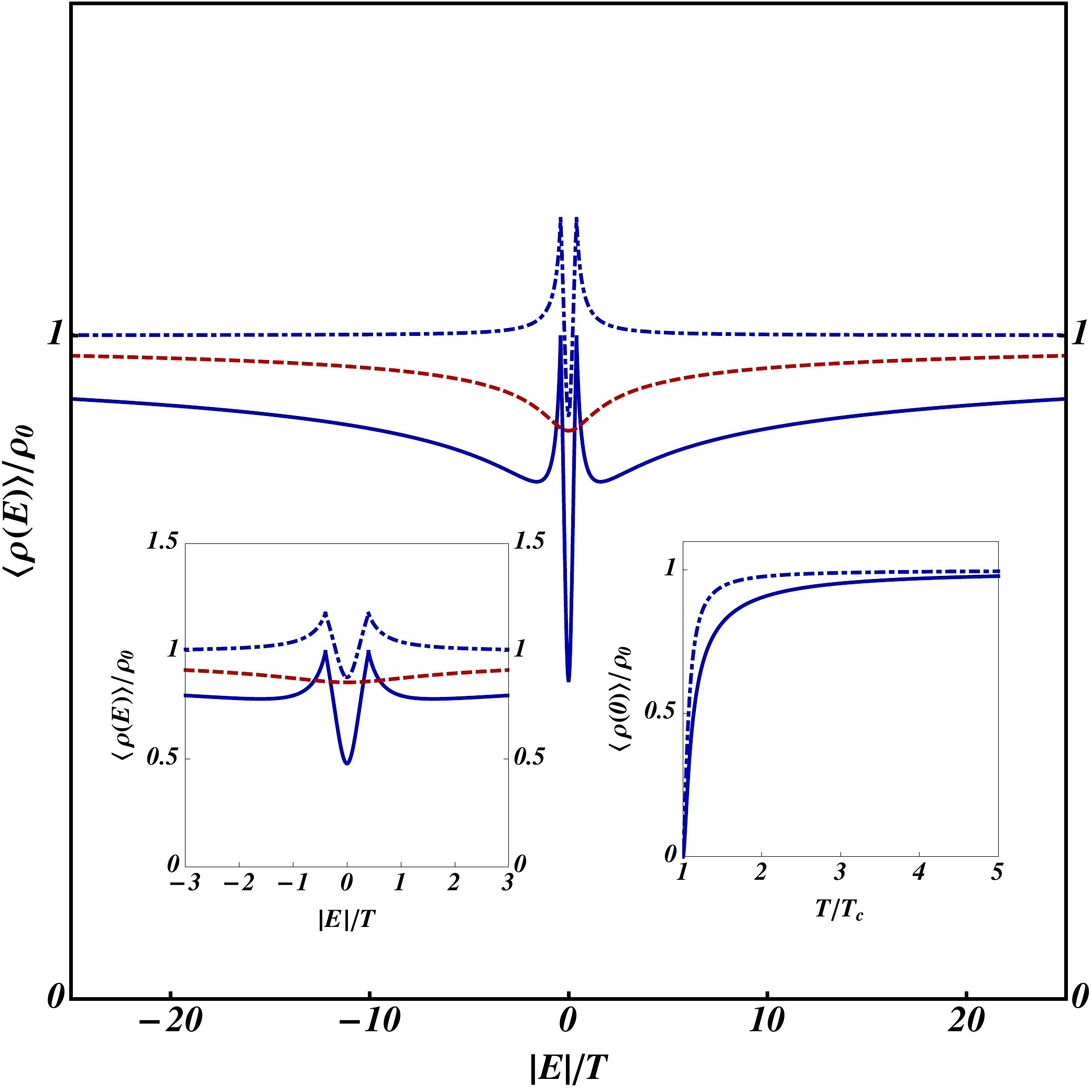}}
	\caption{(Color online) Sketch of the disorder-averaged LDOS $\langle \rho(E) \rangle$ at criticality.
	The solid blue curve corresponds to a temperature $T$ slightly above $T_c^\ast$, while the dashed red curve is plotted for $T\gg T_c^\ast$, see circles of the corresponding colors in Fig.~\protect\ref{Fig:Fig3}a.
	The dot-dashed blue curve is plotted for a temperature $T$ slightly above $T_c^\ast$ according to expressions which ignore renormalization (the factor $Z_\ast^{1/2}(T)$).
	The left inset: enlarged view of dependence of  $\langle \rho(E) \rangle$ on $|E|/T$. The right inset: dependence of  the disorder-averaged LDOS at $E=0$ on temperature.   	See text and Eqs. \eqref{eq:1:e11:d} and \eqref{eq:1:e11a:d}.}
	\label{Fig:Fig4}
\end{figure}

We start our analysis of the disorder-averaged LDOS at the criticality, $\delta_\xi\ll T_c^*$.
At temperatures higher than the superconducting transition temperature, $T\gg T_c^*$,
we can solve Eqs. \eqref{eq:AT:e4.1} and \eqref{eq:AT:LDOS1} for $|\gamma| \ll 1$:
\begin{equation}
\frac{\langle\rho(E)\rangle}{\rho_0} = 1+ \frac{b}{|\Delta_2|} \bigl (\gamma - \gamma_0 \bigr )  .
\end{equation}
Hence we find the following behavior of the disorder-averaged LDOS in the critical region:
\begin{align}
\frac{\langle\rho(E)\rangle}{\rho_0}
= 1-\frac{b}{a}\left (\frac{T_c^*}{|E|}\right )^{|\Delta_2|/d}  + \frac{b}{a}(T_c^*\tau)^{|\Delta_2|/d}
\label{eq:LDOS:crit:HE1}
\end{align}
for $|E| \gg T \gg T_c^*$
and
\begin{align}
\frac{\langle\rho(E)\rangle}{\rho_0}  = 1-\frac{b}{a} \left (\frac{T_c^*}{T}\right )^{|\Delta_2|/d}
+ \frac{b}{a}(T_c^*\tau)^{|\Delta_2|/d}
\label{eq:LDOS:crit:HE2}
\end{align}
for $T\gg |E|\gg T_c^*$. Thus, there is a relatively small depletion of the disorder-averaged LDOS at temperatures $T\gg T_c^*$, see red line in Fig. \ref{Fig:Fig4}.

When temperature is close to the superconducting transition temperature, $T-T_c^* \ll T_c^*$,
the energy dependence of the disorder-averaged LDOS is described by expressions
analogous to Eqs. \eqref{eq:1:e11a}-\eqref{eq:1:e11c}:
\begin{eqnarray}
\frac{\langle \rho(E) \rangle}{\rho_0} &=& Z^{1/2}_*(T)
\notag
\\
&\times&
\begin{cases}
1 - c_d t_\ast \bigl [ T_c^* \tau_{GL}\bigr ]^\frac{6-d}{2}\!\!,  \quad & |E| \ll \tau_{GL}^{-1} , \\
\displaystyle 1 + \tilde{c}_d t_\ast \left [\frac{T_c^*}{|E|}\right ]^\frac{6-d}{2}\!\!\!,\quad  & \tau_{GL}^{-1} \ll |E| \ll T^*_c .\\
\end{cases} \notag \\
\label{eq:1:e11:d}
\end{eqnarray}
The numerical coefficients $c_d$ and $\tilde{c}_d$ are of the order unity
and are given in Appendix \ref{Sec:App:LDOS}. The function
\begin{equation}
Z^{1/2}_*(T) = \left [\frac{\Delta_2+a\gamma_0}{\Delta_2+a\gamma(L_T)} \right ]^{b/a}
\end{equation}
can be estimated for temperatures close to $T_c^*$ as
\begin{equation}
Z^{1/2}_*(T) = \left ( \frac{8 a}{\pi |\Delta_2|} T_c^*\tau_{GL}\right )^{-b/a} .
\end{equation}
At energies $\tau^{-1}\gg |E|\gg T_c^*$, the disorder-averaged LDOS is given by
\begin{gather}
\frac{\langle \rho(E) \rangle}{\rho_0} = 1  - \frac{b}{a}\left (\frac{T_c^*}{|E|}\right )^\frac{|\Delta_2|}{d}
+ \tilde{c}_d t_\ast \left (\frac{T_c^*}{|E|}\right )^\frac{6-d}{2}
.
\label{eq:1:e11a:d}
\end{gather}
The fluctuation correction [the last term on the r.h.s. of Eq.~\eqref{eq:1:e11a:d}] is small
in comparison with the RG correction [the second term on the r.h.s. of Eq. \eqref{eq:1:e11a:d}]
provided the following inequality holds: $2|\Delta_2|<d(6-d)$. The latter is true for $d=3$.
The LDOS at a temperature $T$ close to $T_c$ is shown by the full blue line in Fig. \ref{Fig:Fig4}.
The curves shows a strong suppression of LDOS near zero energy as well as clear precursors of coherence peaks at
$|E| \sim \tau_{GL}^{-1}$.

We note that, strictly speaking, our results \eqref{eq:1:e11:d} are valid under the condition that
the critical resistance is small $t_\ast \ll 1$. Only in this case, there is a wide range of
applicability for Eq. \eqref{eq:1:e11:d}, $1\ll T_c^* \tau_{GL} \ll t_\ast^{-2/(6-d)}$.
For $t_\ast \sim 1$, the results indicate that already at $T_c^*\tau_{GL} \sim 1$ there
is strong suppression of the disorder-averaged LDOS.
In this case one needs to sum up higher-order contributions coming from the superconducting fluctuations.

The scaling behavior of the $q$-th moment of LDOS can be described by the following RG equation:
\begin{equation}
\frac{d\ln m_q}{dy} = -\Delta_q + b_q \gamma .
\label{eq:mq:rg:fp}
\end{equation}
The perturbative result \eqref{eqmqRG} implies $b_q=q(1-q)t^2<0$ for $q > 1$.
We remind the reader that at finite energy and temperature the RG flow, Eq. \eqref{eq:mq:rg:fp},
stops at the length scale $\mathcal{L}$. At larger scales, the interaction correction disappears
from Eq. \eqref{eq:mq:rg:fp} and the scaling of $m_q$ is exactly the same as without interaction
up to the length scale $L_\phi \sim \tau_{\phi}^{1/d}$ induced by the
interaction (see Appendix \ref{app:sec:2}).
We expect a power-law dependence of $\tau_\phi$ on energy and temperature,
$\tau_\phi \sim (\max\{|E|,T\})^{-p}$.

The solution of Eqs. \eqref{eq:AT:e4.1} and \eqref{eq:mq:rg:fp} provides, in particular,
the following result for the temperature/energy dependence of $m_q$:
\begin{equation}
m_q = \left ( \frac{\mathcal{L}}{L_\phi}\right )^{\Delta_q}\left ( \frac{\gamma(\mathcal{L})}{\gamma_0}\right )^{\frac{\Delta_q}{\Delta_2}}
\left ( \frac{\Delta_2+a \gamma_0}{\Delta_2+a \gamma(\mathcal{L})}\right )^{x_q} ,
\end{equation}
where we introduce the exponent $x_q=b_q/a+ \Delta_q/\Delta_2$.
Here we have taken into account that typically $L_\phi \gg \mathcal{L}$.
Hence, at $T\gg T_c^*$ we obtain the following results for the moments of LDOS:
\begin{equation}
\frac{\langle \rho^q(E)\rangle}{\langle \rho(E)\rangle^q}
= \left ( \frac{\tau}{\tau_\phi} \right )^\frac{\Delta_q}{d}
\left [1 -x_q \left ( \frac{T_c^*}{|E|} \right )^\frac{|\Delta_2|}{d} +x_q  \left ( T_c^*\tau \right )^\frac{|\Delta_2|}{d} \right ]
\label{mq:s:1}
\end{equation}
for $|E|\gg T \gg T_c^*$ and
\begin{equation}
\frac{\langle \rho^q(E)\rangle}{\langle \rho(E)\rangle^q}
= \left (  \frac{\tau}{\tau_\phi}  \right )^\frac{\Delta_q}{d}
\left [1 - x_q \left ( \frac{T_c^*}{T} \right )^\frac{|\Delta_2|}{d} +x_q  \left ( T_c^*\tau \right )^\frac{|\Delta_2|}{d} \right ]
\label{mq:s:2}
\end{equation}
for $T \gg |E| \gg T_c^*$.
At temperatures close to the superconducting transition temperature, $T-T_c^*\ll T_c^*$,
the $q$-th moment of the LDOS at energies $|E| \gg T\approx T_c^*$ is given by Eq. \eqref{mq:s:1}.
At energies $|E|\ll T\approx T_c^*$, the $q$-th moment of the LDOS can be estimated as
\begin{equation}
\frac{\langle \rho^q(E)\rangle}{\langle \rho(E)\rangle^q}
= \left (  \frac{\tau}{\tau^*_\phi} \right )^{\frac{\Delta_q}{d}}
\left ( \frac{8 a}{\pi |\Delta_2|} T_c^* \tau_{GL}\right )^{-\frac{b_q}{a}} .
\label{107}
\end{equation}
Here $\tau^*_\phi$ is the dephasing time at $|E|\sim T \sim T_c^*$.
We note that the moments of the normalized LDOS  $\rho(E)/\langle \rho(E)\rangle$
are strongly enhanced at $|E|\sim T\sim T_c^*$ for $b_q<0$.
Provided a stronger condition is fulfilled, $b_q+qb<0$, the moments $\langle\rho^q\rangle(E)$ are large at $|E|\sim T\sim T_c^*$ not only in comparison with $\langle \rho(E)\rangle^q$ but even in comparison with the bare value $\rho_0^q$.

\subsection{Off criticality: Metallic / superconducting side}
\label{s5b}

Let us now consider the metallic side of the Anderson transition, $t_0<t_*$.  In this case the system is a superconductor for temperatures below $T_c(\xi)$ given by Eq.~\eqref{eq:AT:e4.7}. We will assume that the system is off criticality, in the sense that $\delta_\xi \gg T_c^*$, in which case $T_c(\xi)\ll T_c^*$.
At high temperatures, $T\gg T_c(\xi)$, the disorder-averaged LDOS reads as
\begin{equation}
\frac{\langle\rho(E)\rangle}{\rho_0}
=1+  \frac{b}{|\Delta_2|} \gamma(\xi) \left ( \frac{\delta_\xi}{\max\{|E|,T,\delta_\xi\}}\right )^\frac{|\Delta_2|}{d}  .
\label{eq:LDOS:CR1:s}
\end{equation}
This result is the solution of RG equation \eqref{eq:AT:LDOS1} taken at the length scale $\min\{\xi,\mathcal{L}\}$.
We note that the attraction interaction at the scale $\xi$ can be expressed via the
superconducting transition temperature: $\gamma(\xi) = 1/\ln[T_c(\xi)/\delta_\xi]$.

For temperatures close to $T_c(\xi)$, $T-T_c(\xi) \ll T_c(\xi)$, the disorder-averaged LDOS becomes
\begin{equation}
\frac{\langle\rho(E)\rangle}{\rho_0}  =1+  \frac{b}{|\Delta_2|} \gamma(\xi) \left ( \frac{\delta_\xi}{|E|}\right )^\frac{|\Delta_2|}{d}  + \tilde{c}_d t_\ast \left (\frac{T_c(\xi)}{|E|}\right )^\frac{6-d}{2}.
\label{eq:LDOS:CR1:s1}
\end{equation}
for $|E|\gg \delta_\xi$ and
\begin{equation}
\frac{\langle\rho(E)\rangle}{\rho_0}  =1+  \frac{b}{|\Delta_2|} \gamma(\xi)  .
\label{eq:LDOS:CR1:s2}
\end{equation}
for $|E|\ll \delta_\xi$. We note that the energy dependence of the disorder-averaged LDOS for $|E|\gg \delta_\xi$
is the same as at the criticality (cf. Eqs. \eqref{eq:LDOS:crit:HE1}, \eqref{eq:LDOS:CR1:s1} and \eqref{eq:1:e11a:d}).
There is no disorder-induced fluctuation corrections like in Eq. \eqref{eq:1:e11:d}
at energies $|E|\ll \delta_\xi$ since $t(\xi)=0$.

The dependence of the $q$-th moment of LDOS on energy and temperature at $T>T_c(\xi)$
is determined by the solution of RG equation \eqref{eq:mq:rg:fp}:
\begin{align}
\frac{\langle \rho^q(E)\rangle}{\langle \rho(E)\rangle^q}  & = \left ( \frac{\gamma(\xi)}{\gamma_0}\right )^\frac{\Delta_q}{\Delta_2} \Bigl (\min\{1,{\delta_\xi}\tau_\phi\} \Bigr )^\frac{|\Delta_q|}{d} \notag  \\
& \times
\left [1 - \frac{x_q a}{\Delta_2} \gamma(\xi)
\left ( \min\left \{1,\frac{\delta_\xi}{|E|},\frac{\delta_\xi}{T}\right \}\right )^\frac{|\Delta_2|}{d} \right ] .
\label{115}
\end{align}
We note that the dephasing rate here is affected by the superconducting fluctuations, see Appendix \ref{app:sec:2}.

\subsection{Off criticality: Insulating side}
\label{s5c}

Finally, let us consider the insulating phase of the Anderson transition, $t_0> t_\ast$. We will assume that we are sufficiently far from criticality, $\delta_\xi \gtrsim T_c^*$, in which case the system remains an  insulator in the presence of attracting interaction (i.e., we are on the insulating side of SIT).
For the sake of simplicity, we consider the zero-temperature regime, $T=0$.
At high energies, $\tau^{-1}\gg |E|\gg \delta_\xi$, the disorder-averaged LDOS
can be found from Eqs. \eqref{eq:AT:e4.1} and \eqref{eq:AT:LDOS1}:
\begin{equation}
\frac{\langle\rho(E)\rangle}{\rho_0}  =1+  \frac{b}{|\Delta_2|} \gamma(\xi) \left ( \frac{\delta_\xi}{|E|}\right )^\frac{|\Delta_2|}{d}  .
\label{eq:LDOS:CR1}
\end{equation}
We emphasize that $|\gamma(\xi)| \ll 1$  at $\delta_\xi \gg T_c^*$.
We note that the result \eqref{eq:LDOS:CR1} is the same as Eq. \eqref{eq:LDOS:crit:HE1}
in which we neglect the last term on the r.h.s. in comparison with the second one.
Therefore, at high energies $|E| \gg \delta_\xi$ the energy dependence of the
disorder-averaged LDOS is the same as in the critical region, $\delta_\xi \ll T_c^*$.

At $|E|\ll \delta_\xi$, one can still use Eq. \eqref{rhores}, but
taking into account the insulator-type behavior of the conductivity (cf. Ref.~[\onlinecite{Burmistrov2014}]).
Then we obtain (see Appendix \ref{Sec:App:LDOS}):
 \begin{equation}
\frac{\langle\rho(E)\rangle}{\rho_0}  =1 - a_1 |\gamma(\xi)| \left ( \ln \frac{\delta_\xi}{|E|}\right )^{d+2}
- a_2 \gamma^2(\xi) \frac{\delta_\xi}{|E|}.
\label{eq:rho:ins1}
\end{equation}
Here $a_1$ and $a_2$ are positive constants.
The last term on the r.h.s. of Eq. \eqref{eq:rho:ins1} is small at $|E|\sim \delta_\xi$.
With further lowering energy,
it becomes larger than the second term
in the energy interval $|\gamma(\xi)| \Delta_P\ll |E| \ll \Delta_P$, where
\begin{equation}
\Delta_P \sim |\gamma(\xi)| \delta_\xi \propto \xi^{-d-\Delta_2}.
\label{DeltaP}
\end{equation}
We emphasize that the energy scale $\Delta_P$ coincides with the so-called
pseudogap energy scale introduced in Ref. [\onlinecite{FeigelmanCuevas2010}].

\begin{figure}[t]
	\centerline{\includegraphics[width=0.4\textwidth]{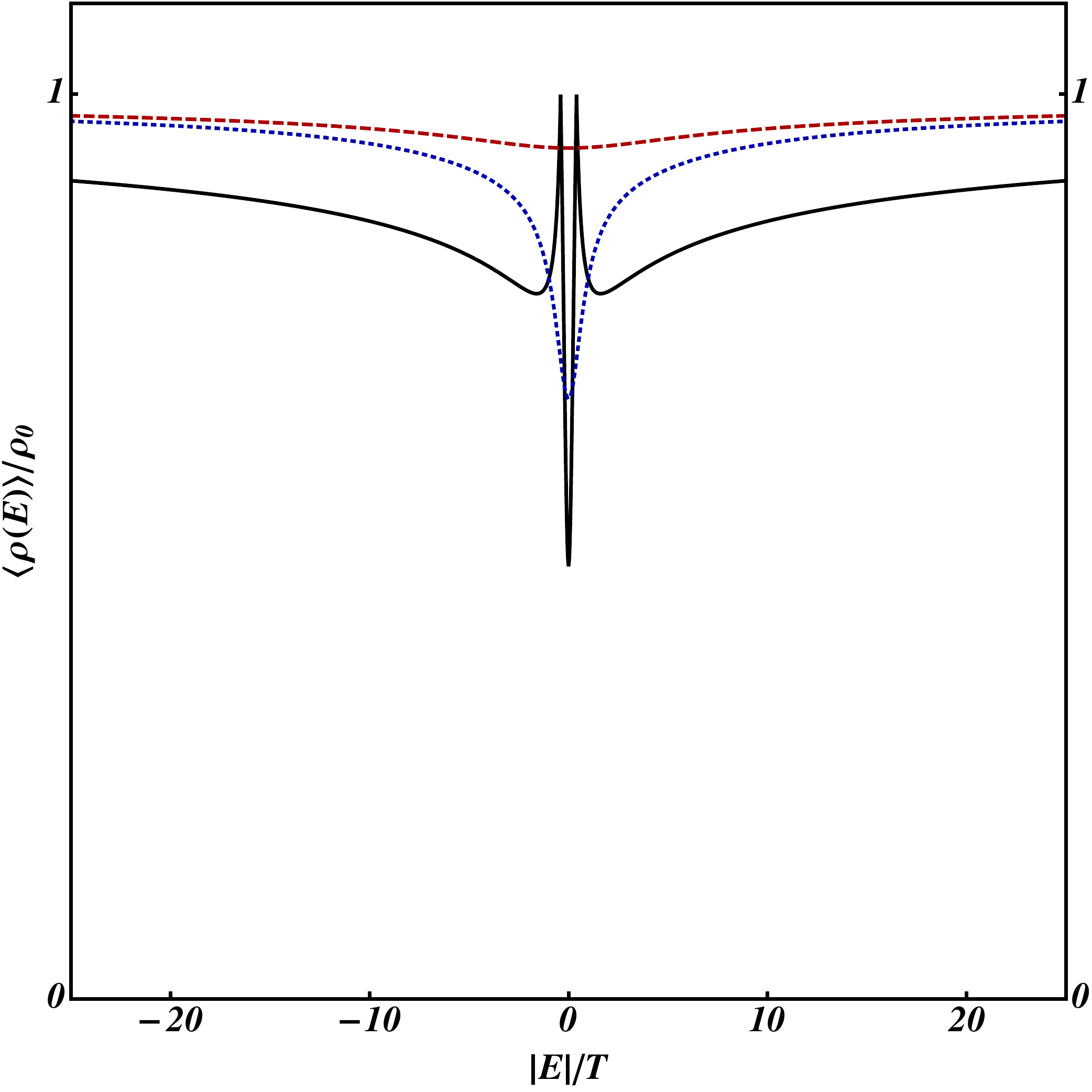}}
	\caption{(Color online) Disorder-average LDOS $\langle \rho(E) \rangle$ across  the SIT.
	All three curves corresponds to the same temperature $T$, which is assumed to be slightly exceeding the superconducting transition temperature $T_c^\ast$ at the Anderson transition, see triangles of the corresponding colors in Fig.~\ref{Fig:Fig3}a.
	The dashed red curve corresponds to the metallic side of the Anderson transition, $t_0<t_*$, where the interacting system is a superconductor at low temperatures  [Sec.~\ref{s5b}, Eqs.~\eqref{eq:LDOS:CR1:s1} and \eqref{eq:LDOS:CR1:s2}]. The solid black curve is the LDOS at criticality, $t_0=t_*$ [Sec.~\ref{s5a}, Eqs. \eqref{eq:1:e11:d} and \eqref{eq:1:e11a:d}], and  the dotted blue curve is plotted for the insulating phase $t_0>t_*$ [Sec.~\ref{s5c}, Eq.~\eqref{eq:rho:ins1})].
	}
	\label{Fig:Fig5}
\end{figure}

The behavior  of the disorder-averaged LDOS $\langle \rho(E) \rangle$ across  the SIT is shown in Fig. \ref{Fig:Fig5}. All three curves in this figure corresponds to the same temperature $T$, which is assumed to be slightly exceeding the superconducting transition temperature $T_c^*$ at the Anderson transition, see triangles of the corresponding colors in Fig.~\ref{Fig:Fig3}a. The red curve shows a weak depletion on the metallic side of the Anderson transition, $t_0<t_*$ (Sec.~\ref{s5b}). Note that for this curve the temperature $T$ is much larger than the relevant superconducting temperature $T_c(\xi)$; the depletion becomes more pronounced when one reduces the temperature, approaching $T_c(\xi)$. The black curve corresponds to criticality, $t_0=t_*$ (Sec.~\ref{s5a}). It shows a much stronger suppression of LDOS around zero energy, accompanied by clear precursors of coherence peaks (as was already shown by solid blue curve in Fig.~\ref{Fig:Fig4}. Finally, the blue curve in Fig. \ref{Fig:Fig5} illustrates a strong ``pseudogap''
arising on the insulating side of the transition.

For high energies, $\tau^{-1}\gg |E|\gg \delta_\xi$, the fluctuations of LDOS are
controlled by the noninteracting fixed point at $t=t_\ast$.
As in the critical region, the moments of LDOS are determined by the
corresponding multifractal exponents $\Delta_q$:
\begin{align}
\frac{\langle \rho^q(E)\rangle}{\langle \rho(E)\rangle^q}
& = \left ( \frac{\gamma(\xi)}{\gamma_0}\right )^\frac{\Delta_q}{\Delta_2}
\left (\frac{\delta_\xi}{|E|} \right )^\frac{|\Delta_q|}{d} \notag  \\
& \times
\left [1 - \frac{x_q a}{\Delta_2} \gamma(\xi)
\left ( \frac{\delta_\xi}{|E|}\right )^\frac{|\Delta_2|}{d} \right ].
\label{110}
\end{align}
We note that the last term in the square brackets on the r.h.s. of Eq. \eqref{eq:mq:rg:fp}
is much smaller than unity since the attractive interaction at the scale of localization length
is weak, $|\gamma(\xi)|\ll 1$. When energy is below the level spacing in the localization volume, $|E| \ll \delta_\xi$,
the LDOS shows the multifractal behavior up to the scale $\xi$ and, then,
insulator-like fluctuations up to the system size $L$:
\begin{equation}
\frac{\langle \rho^q(E)\rangle}{\langle \rho(E)\rangle^q}  =
\left ( \frac{\gamma(\xi)}{\gamma_0}\right )^\frac{\Delta_q}{\Delta_2} \left (\frac{L}{\xi}\right )^{d(q-1)} .
\label{111}
\end{equation}
At finite temperatures $T\ll |E|$, the insulator-like fluctuations are regularized
by the temperature-induced dephasing length $L_{\phi T}$.

\section{Summary and Conclusions}
\label{s6}

In conclusion, we have developed the theory of local density of states and its mesoscopic fluctuations near the transition to the superconducting state.  Specifically, we considered systems on the superconducting side of SIT but at temperatures above $T_c$, as well as systems on the insulating side of SIT.  We have employed the non-linear sigma-model
formalism and constructed the operators that describe the moments of the local density of states.
Our strategy has combined the two steps: (i) renormalization of these operators as well as the coupling constants
of the action followed by (ii) including the superconducting fluctuations arising from real processes.
In view of the length of the paper, we find it appropriate to list here the main results obtained in Sections \ref{s3} -- \ref{s5}, with the references to corresponding equations.

\begin{enumerate}
\item In Sec. \ref{s3}, we have performed the two-loop renormalization of the moments of LDOS in the presence of interactions
in the singlet and triplet particle-hole channels, as well as in the Cooper channel. The
zeta-function governing the renormalization of the $q$th moment of the LDOS is given by Eq.~(\ref{eqmqRG}).

\item  In Sec. \ref{s4}, we have used these two-loop results to study the scaling behavior of
the disorder-averaged LDOS and its moments in 2D disordered films near the finite-temperature superconducting transition.
For short-ranged interactions, the evolution of the average LDOS $\langle\rho(E)\rangle$ with increasing energy $E$
is given by Eqs.~(\ref{eq:1:e11a})-(\ref{eq:1:e11c}) and illustrated in Fig.~\ref{Fig:Fig1}. We have found that
the combined effect of renormalization and superconducting fluctuations progressively depletes the LDOS with lowering energy $|E|$ and suppresses the ``coherence peaks''.

\item  The scaling of the moments of the LDOS for short-range interactions is given by Eqs.~\eqref{eq:1:e12a} and \eqref{eq:1:e12b}. 
  The LDOS fluctuations are particularly strong for weak initial couplings satisfying $|\gamma_0|\ll t_0\ll \sqrt{|\gamma_0|}$, which is the range of parameters where the enhancement of $T_c$ by multifractality takes place. A representative curve characterizing LDOS fluctuations for intermediate initial couplings shown in   Fig. \ref{Fig:Fig1fl} demonstrates that the fluctuations are rather strong also in this regime.

\item Further, in Sec. \ref{subsec:2DCoulomb}, we have analyzed the LDOS in 2D superconducting films with long-range Coulomb
repulsion. The superconducting transition temperature is then suppressed as compared to the clean case.
The evolution of the average LDOS with energy is described by Eqs.~(\ref{eq:t:c:3a})-(\ref{eq:t:c:3c}), see Fig.~\ref{Fig:Fig2},
and is governed by the interplay of the Coulomb-induced zero-bias anomaly and superconducting fluctuations.
The overall behavior of the LDOS is similar to the case of the short-range interactions, cf. Figs. \ref{Fig:Fig1} and \ref{Fig:Fig2}. Mesoscopic fluctuations of the LDOS, Eqs.~(\ref{eq:t:c:2a}) and (\ref{eq:t:c:2b}), are found to be suppressed by the Coulomb repulsion.
A representative curve characterizing LDOS fluctuations for intermediate initial couplings is shown in   Fig. \ref{Fig:Fig2fl}. 
The fluctuations are substantially weaker than for the case of short-range interaction with comparable bare couplings (Fig. \ref{Fig:Fig1fl}) but remain quite sizeable.

\item In Sec. \ref{s5}, we have studied the LDOS near the superconducting transition in a system with weak short-ranged interactions which, in the absence of interactions, would be close to the Anderson metal-insulator transition. On the superconducting side, the average LDOS is described by
Eqs. (\ref{eq:LDOS:CR1:s})-(\ref{eq:LDOS:CR1:s2}), and the scaling of its $q$-th moment is given  by Eq. (\ref{115}). The average LDOS  decreases with lowering of energy at high energies, $|E|\gg \delta_\xi$,  and saturates at $|E|\ll \delta_\xi$, see Fig. \ref{Fig:Fig5}. We neglected suppression of $\langle\rho(E)\rangle$ arising from the ballistic scales. At criticality, the average LDOS is given by Eqs.~(\ref{eq:1:e11:d}) and
(\ref{eq:1:e11a:d}) and is strongly depleted around zero energy by superconducting fluctuations, as illustrated in Figs.~\ref{Fig:Fig4} and \ref{Fig:Fig5}. The corresponding LDOS fluctuations are described by Eqs.~(\ref{mq:s:1})-(\ref{107}) and are strong for energies and temperatures of the order of critical temperature $T_c^*$.

\item On the insulating side, the average LDOS is also strongly depeleted, see Eqs. (\ref{eq:LDOS:CR1}) and (\ref{eq:rho:ins1}) and Fig. \ref{Fig:Fig5}. The energy scale $\Delta_P$, Eq.~(\ref{DeltaP}),
which emerges as the characteristic scale in these formulas for  $\langle\rho(E)\rangle$,
resembling the pseudogap scale introduced in Ref. [\onlinecite{FeigelmanCuevas2010}].
The LDOS fluctuations on the insulating side
at high and low energies (compared to the level spacing in the localization volume) are given by Eqs. (\ref{110}) and (\ref{111}), respectively.

\end{enumerate}

Let us now summarize on a qualitative level the most salient of these findings.

\begin{enumerate}

\item
We have observed a strong depletion of LDOS  in two regimes:
(i) on the superconducting side of SIT, above $T_c$, and
(ii) on the insulating side of SIT.

\item This depletion arises from a combination of two mechanisms: (i) renormalization effects that are operative at higher energies $E\gtrsim T$, and (ii) real processes due to superconducting fluctuations that are operative at lower energies $E\lesssim T$.   The renormalization effects are governed by attractive interaction for systems with short-range interaction or by Coulomb interaction when it is present. Remarkably, the resulting depletion of LDOS is qualitatively similar in these two cases.

\item The interplay of renormalization effects and of superconducting fluctuations tends to suppress the precursors of coherence peaks.

\item
A substantial depletion of LDOS remains observable for temperatures much exceeding (by factor $\sim 10$) the superconducting transition temperature $T_c$.

\item In a model with short-range interaction, multifractality leads to strong mesoscopic fluctuations of LDOS, which should be observable as point-to-point fluctuations when the surface of a system is scanned in an STM experiment. The Coulomb interaction reduces the magnitude of the mesoscopic fluctuations. However, also in a model with Coulomb interaction, the fluctuations become strong when the systems approaches the SIT.

\end{enumerate}

Our findings compare well with experimental observations of depletion of LDOS
 and of its large point-to-point fluctuations in the metallic and insulating phases near SIT in
TiN, InO, and NbN films [\onlinecite{Sacepe08,Sacepe10,Sacepe11,Sherman14,Mondal11,Noat13},\onlinecite{note-samuely}].
Let us emphasize that our results have been obtained entirely within the sigma-model formalism (usually referred to as a ``fermionic approach'') for a macroscopically homogeneous system. All the observed effects are thus intrinsic properties of this problem and do not require any additional assumptions, such as the presence of macroscopic inhomogeneities (``granularity'').

\begin{acknowledgments}

We thank M.V. Feigel'man, M.A. Skvortsov and K.S. Tikhonov for useful discussions. The work was supported by the DFG and by the Russian Science Foundation under the grant No. 14-42-00044.

\end{acknowledgments}

\appendix

\begin{widetext}

 \section{Two-loop contribution to the LDOS correlation function $K_2$}
 \label{Sec:App:2loop}

In this Appendix we present technical details of the calculation of the irreducible two-point LDOS correlation function $K_2$.
 We start with the evaluation of the two-loop contribution from $P^{\alpha_1\alpha_2}_2(i\varepsilon_{n_1},i\varepsilon_{n_3})$.
 In the two-loop approximation, one needs to take into account only the terms with four $W$:
\begin{gather}
[P_2^{\alpha_1\alpha_2}]^{(2)}(i\varepsilon_{n_1},i\varepsilon_{n_3}) = \frac{1}{4} \sum_{n_6n_8}\sum_{\beta_1\beta_2} \Bigl [ \langle \langle \spp \bigl [
w^{\alpha_1\beta_1}_{n_1n_6}\bar{w}^{\beta_1\alpha_1}_{n_6n_1} \bigr ]
\cdot \spp \bigl [
w^{\alpha_2\beta_2}_{n_3n_8}\bar{w}^{\beta_2\alpha_2}_{n_8n_3}  \bigr ]
\rangle \rangle - 2
\spp \bigl\langle
 w^{\alpha_1\beta_1}_{n_1n6}\bar{w}^{\beta_1\alpha_2}_{n_6n_3}
w^{\alpha_2\beta_2}_{n_3n_8}\bar{w}^{\beta_2\alpha_1}_{n_8n_1}
\bigr \rangle \Bigr ] .
\end{gather}
By using Wick's theorem and Eqs.~\eqref{eq:prop:PH} - \eqref{eq:prop:PPS}, we find
\begin{gather}
[P_2^{\alpha_1\alpha_2}]^{(2)}(i\varepsilon_{n_1},i\varepsilon_{n_3})  =
\frac{2^{12}\pi T}{g^3}\sum_{j=0}^3 \Gamma_j\int_{q,p}  \sum_{\omega_n>\varepsilon_{n_3}} \mathcal{D}_p(i\omega_n)
\mathcal{D}^{(j)}_p(i\omega_n)  \Bigr [ \mathcal{D}_q(i\omega_n+i\Omega^\varepsilon_{13}) + \mathcal{C}_q(i\omega_n+i\Omega^\varepsilon_{13}) \Bigl ]
\notag \\
+
\frac{2^{13}\pi T Z_\omega}{g^3}\int_{q,p}\sum_{\omega_n<\varepsilon_{n_3}}
\mathcal{C}^2_p(2i\varepsilon_{n_3}-i\omega_n) \mathcal{L}_p(i\omega_n)
\Bigr [ \mathcal{D}_q(i\mathcal{E}_{13}-i\omega_n) +  \mathcal{C}_q(i\mathcal{E}_{13}-i\omega_n) \Bigl ]
+ (\varepsilon_{n_1} \leftrightarrow \varepsilon_{n_3})
.
\label{eq2loopK2_P2++1}
\end{gather}
Performing the analytical continuation to real frequencies, $i\varepsilon_{n_1} \to E+i0^+$, $i\varepsilon_{n_3} \to E^\prime+i0^+$, we obtain
\begin{align}
[P_2^{\alpha_1\alpha_2}]^{RR (2)}(E,E^\prime) & =
\frac{2^{11}}{ig^3} \sum_{j=0}^3 \Gamma_j \int_{q,p,\omega} \mathcal{F}_{\omega-E^\prime} \mathcal{D}^R_p(\omega)
\mathcal{D}^{(j) R}_p(\omega) \mathcal{D}^R_q(\omega+\Omega)
\notag \\
& +
\frac{2^{12}Z_\omega}{ig^3}\int_{q,p,\omega}
\mathcal{C}^{R2}_p(2E-\omega)
\mathcal{C}^R_q(\mathcal{E}-\omega)
\bigl [
 \mathcal{L}^K_p(\omega)+\mathcal{F}_{E-\omega} \mathcal{L}^R_p(\omega)
 \bigr ]
  + (E \leftrightarrow E^\prime).
\label{eq2loopK2_P2++11}
\end{align}
where $\mathcal{E}=E+E^\prime$.
Here we have taken into account that diffuson $\mathcal{D}^R_q(\omega)$ and
cooperon $\mathcal{C}^R_q(\omega)$ propagators are the same.
Setting $E=E^\prime=T=0$, we find in  $d=2+\epsilon$ dimensions (see details for evaluation of the integrals in Ref.~[\onlinecite{Burmistrov2015a}])
\begin{gather}
[P_2^{\alpha_1\alpha_2}]^{RR (2)} \to 16 \frac{t^2\,h^{2\epsilon}}{\epsilon^2} \Bigl [ 2\gamma_c+ \sum_{j=0}^3 \ln (1+\gamma_j)
 -\frac{\epsilon}{4} \sum_{j=0}^3 \ln^2(1+\gamma_j) \Bigr ] + O(\epsilon).
\label{eq2loopK2_P2++2}
\end{gather}
We note that the result \eqref{eq2loopK2_P2++2} is of the first order in $\gamma_c$.
This occurs since the terms of the first order in the fluctuation propagator exist in Eq. \eqref{eq2loopK2_P2++11}.

Next, the two-loop contribution to $P^{\alpha_1\alpha_2}_2(i\varepsilon_{n_1},i\varepsilon_{n_2})$ can be written as follows
\begin{gather}
[P^{\alpha_1\alpha_2}_2]^{(2)}(i\varepsilon_{n_1},i\varepsilon_{n_2}) =
-\frac{1}{4} \sum_{n_5n_6}\sum_{\beta_1\beta_2} \langle \langle \spp \bigl [
w^{\alpha_1\beta_1}_{n_1n_6}\bar{w}^{\beta_1\alpha_1}_{n_6n_1} \bigr ]
\cdot
\spp \bigl [\bar{w}^{\alpha_2\beta_2}_{n_2n_5}{w}^{\beta_2\alpha_2}_{n_5n_2}  \bigr ]
\rangle \rangle
\notag
\\
- 2 \Bigl \langle \spp \bigl [w^{\alpha_1\alpha_2}_{n_1n_2} \bar{w}^{\alpha_2\alpha_1}_{n_2n_1}\bigr ]
\Bigl [ S^{(4)}_\sigma+S^{(4)}_{\rm int}+\frac{1}{2} \left (S^{(3)}_{\rm int}\right )^2\Bigr ] \Bigr \rangle .
\label{eqP2+-0}
\end{gather}
Here the term
\begin{align}
S^{(4)}_\sigma   = & -\frac{g}{128} \int_{q_j} \delta\left (\sum_{j=0}^3\bm{q_j}\right )
\sum_{\beta_1\beta_2\beta_3\beta_4}\sum_{n_5n_6n_7n_8}
\spp \Bigl [ w^{\beta_1\beta_2}_{n_5n_6}(\bm{q_0}) \bar{w}^{\beta_2\beta_3}_{n_6n_7}(\bm{q_1})
w^{\beta_3\beta_4}_{n_7n_8}(\bm{q_2}) \bar{w}^{\beta_4\beta_1}_{n_8n_5}(\bm{q_3})\Bigl ]
\notag \\
& \times
 \Bigl [ 2h^2+ \frac{16 z}{g} (\Omega^\varepsilon_{56}+\Omega^\varepsilon_{78})-(\bm{q_0}+\bm{q_1})(\bm{q_2}+\bm{q_3})
 -(\bm{q_0}+\bm{q_3})(\bm{q_1}+\bm{q_2}) \Bigr ] ,
\end{align}
appears in the expansion of $S_\sigma$ and the regulator term \eqref{SsGenFull} to the forth order in $W$.
The expansion of the interaction term $S_{\rm int}$ results in the following third- and forth-order terms,
\begin{align}
S^{(3)}_{\rm int}  & = \frac{\pi T}{4} \sum_{r=0,3}\sum_{j=0}^3 \Gamma_j \sum_{\alpha,n} \int d\bm{r} \Tr I^{\alpha}_{n} t_{rj} W
\Tr I^{\alpha}_{-n} t_{rj}\Lambda W^2
+\frac{\pi T}{4}  \Gamma_c \sum_{\alpha,n} \sum_{r=1,2}  \int d\bm{r} \Tr \bigl [ t_{r0} L_n^\alpha W \bigr ]
 \Tr \bigl [ t_{r0} L_n^\alpha \Lambda W^2 \bigr ]
,
\end{align}
\begin{align}
S^{(4)}_{\rm int} & = -\frac{\pi T}{16} \sum_{r=0,3}\sum_{j=0}^3 \Gamma_j \sum_{\alpha,n} \int d\bm{r} \Tr I^{\alpha}_{n} t_{rj} \Lambda W^2
\Tr I^{\alpha}_{-n} t_{rj}\Lambda W^2
-\frac{\pi T}{16}  \Gamma_c \sum_{\alpha,n} \sum_{r=1,2}  \int d\bm{r} \left ( \Tr \bigl [ t_{r0} L_n^\alpha \Lambda W^2 \bigr ] \right )^2.
\end{align}
After evaluation of averages in Eq.~\eqref{eqP2+-0}, we find
\begin{align}
& [P_2^{\alpha_1\alpha_2}]^{(2)}(i\varepsilon_{n_1},i\varepsilon_{n_2})
=  - \left ( \frac{16}{g}\right )^2
\left [ \left ( \int_q \mathcal{D}_q(i\Omega^\varepsilon_{12})\right )^2 + \left ( \int_q \mathcal{C}_q(i\Omega^\varepsilon_{12})\right )^2 \right ]
\notag \\
& \hspace{0.5cm}
+ \frac{1-9}{4} \left ( \frac{16}{g}\right )^2 \int_{q,p} \Bigl [ p^2+q^2+h^2+\frac{16 z}{g} \Omega^\varepsilon_{12} \Bigr ]
\mathcal{C}_p(i\Omega^\varepsilon_{12}) \mathcal{D}_q(i\Omega^\varepsilon_{12})\Bigl [\mathcal{D}_q(i\Omega^\varepsilon_{12})+\mathcal{C}_p(i\Omega^\varepsilon_{12}) \Bigr ]
\notag \\
& \hspace{0.5cm}
 - \left ( \frac{64}{g}\right )^2 \sum_{j=0}^3 \frac{\pi T\Gamma_j}{g} \int_{q,p}
 \Bigl \{ \sum_{\omega_n>\varepsilon_{n_1}} + \sum_{\omega_n>-\varepsilon_{n_2}}\Bigr \}
 \Bigl [ \mathcal{D}^2_p(i\Omega^\varepsilon_{12})+\mathcal{C}^2_p(i\Omega^\varepsilon_{12})\Bigr ]
\notag \\
& \hspace{1.5cm}
\times
 \Bigl [ p^2+q^2 +2h^2 +\frac{16 Z_\omega}{g} \bigl ( \Omega^\varepsilon_{12} + \omega_n \bigr ) \Bigr ] \mathcal{D}_q(i\omega_n) \mathcal{D}^{(j)}_q(i\omega_n)
  \notag \\
& \hspace{0.5cm}
   + \left ( \frac{64}{g}\right )^2 \sum_{j=0}^{3} \frac{2\pi T\Gamma_j}{g} \int_{q,p} \sum_{\omega_n>0}
   \Bigl [1 - \frac{16 \Gamma_j \omega_n}{g} \mathcal{D}_{\bm{q}+\bm{p}}^{(j)}(i\omega_{n})  \Bigr ]
\Bigl [ \mathcal{D}^2_q(i\Omega^\varepsilon_{12}) \mathcal{D}_p(i\Omega^\varepsilon_{12}+i\omega_n) + \mathcal{C}^2_q(i\Omega^\varepsilon_{12}) \mathcal{C}_p(i\Omega^\varepsilon_{12}+i\omega_n) \Bigr ]
\notag \\
& \hspace{0.5cm}
- \left ( \frac{64}{g}\right )^2 \frac{2\pi T Z_\omega}{g} \int_{q,p}  \Bigl [ \mathcal{D}^2_p(i\Omega^\varepsilon_{12})+\mathcal{C}^2_p(i\Omega^\varepsilon_{12})\Bigr ] \Bigl \{ \sum_{\omega_n>\varepsilon_{n_1}}
\mathcal{L}_q(2i \varepsilon_{n_1}- i\omega_n)
+\sum_{\omega_n>-\varepsilon_{n_2}}\mathcal{L}_q(i\omega_n+2 i\varepsilon_{n_2})
\Bigr \}
\notag \\
& \hspace{1.5cm}\times
\Bigl [ p^2+q^2 +2h^2 +\frac{16 Z_\omega}{g} \bigl ( \Omega^\varepsilon_{12}  + \omega_n \bigr ) \Bigr ]
\mathcal{C}_q^2( i\omega_n)
\notag \\
& \hspace{0.5cm}
+ \left ( \frac{64}{g}\right )^2 \sum_{j=0}^{3} \frac{\pi T\Gamma_j}{g}
 + \sum_{-\varepsilon_{n_2}>\omega_n>0} \Bigr \}
  \Bigl [1 - \frac{16 \Gamma_j \omega_n}{g} \mathcal{D}_{\bm{q}+\bm{p}}^{(j)}(i\omega_{n})  \Bigr ]
     \Bigl [ \mathcal{D}^2_q(i\Omega^\varepsilon_{12}) \mathcal{D}_p(i\Omega^\varepsilon_{12}-i\omega_n) + \mathcal{C}^2_q(i\Omega^\varepsilon_{12}) \mathcal{C}_p(i\Omega^\varepsilon_{12}-i\omega_n) \Bigr ]
 \notag \\
 & \hspace{0.5cm}  +
  \frac{2^{13}\pi T Z_\omega}{g^3} \int_{q,p}
 \Bigl \{  \sum_{\omega_n>\varepsilon_{n_1}} \mathcal{L}_{\bm{q}+\bm{p}}(i\omega_{n}-i\mathcal{E}_{12})
 +
 \sum_{\omega_n>-\varepsilon_{n_2}}\mathcal{L}_{\bm{q}+\bm{p}}(i\omega_{n}+i\mathcal{E}_{12})
 \Bigr \}
\Bigl [ \mathcal{C}^2_q(i\Omega^\varepsilon_{12}) \mathcal{D}_p(i\omega_n)
+ \mathcal{D}^2_q(i\Omega^\varepsilon_{12}) \mathcal{C}_p(i\omega_n)\Bigr ] .
\label{eq2loopK2_P2+-M}
\end{align}
Performing analytic continuation to the real frequencies, $i\varepsilon_{n_1} \to E+i0^+$, $i\varepsilon_{n_2} \to E^\prime-i0^+$, in Eq. \eqref{eq2loopK2_P2+-M}, we obtain
\begin{align}
& [P_2^{\alpha_1\alpha_2}]^{RA (2)}(E,E^\prime) =
- 2 \left ( \frac{16}{g}\right )^2
 \left ( \int_q \mathcal{D}^R_q(\Omega)\right )^2
- \left (\frac{32}{g}\right )^2 \int_{q,p} \Bigl [ p^2+q^2+h^2-\frac{16 Z_\omega}{g} i \Omega \Bigr ]
\mathcal{D}^R_p(\Omega) \mathcal{D}^{R2}_q(\Omega)
\notag \\
& \hspace{0.5cm}
 - 2 \left ( \frac{32}{g}\right )^2 \sum_{j=0}^3 \frac{\Gamma_j}{i g} \int_{q,p,\omega}
\Bigl [ \mathcal{F}_{\omega-E} + \mathcal{F}_{\omega+E^\prime}\Bigr ]
\Bigl [ p^2+q^2 +2h^2 - \frac{16 Z_\omega}{g} i \bigl ( \Omega + \omega \bigr ) \Bigr ]
\mathcal{D}^{R2}_p(\Omega)
\mathcal{D}^R_q(\omega) \mathcal{D}^{(j)R}_q(\omega)
\notag \\
& \hspace{0.5cm}
 - \left ( \frac{64}{g}\right )^2 \frac{Z_\omega}{ig} \int_{q,p,\omega} \mathcal{C}^{R2}_p(\Omega)
 \Bigl \{
\Bigl [ p^2+q^2 +2h^2 -\frac{16 Z_\omega}{g} i\bigl ( \Omega  + 2 E- \omega \bigr ) \Bigr ]
\mathcal{C}_q^{R2}(2E- \omega) \bigr [ \mathcal{L}^K_q(\omega) +\mathcal{F}_{E-\omega}\mathcal{L}^R_q(\omega) \bigl ]
\notag \\
& \hspace{0.5cm}
+
\Bigl [ p^2+q^2 +2h^2 -\frac{16 Z_\omega}{g} i \bigl ( \Omega+ \omega - 2 E^\prime \bigr ) \Bigr ]
\mathcal{C}_q^{R2}(\omega - 2 E^\prime)\bigr [ \mathcal{L}^K_q(\omega)+ \mathcal{F}_{\omega-E^\prime}\mathcal{L}^A_q(\omega) \bigl ] \Bigr \}
 \notag \\
  & \hspace{0.5cm}
  + \left ( \frac{64}{g}\right )^2 \sum_{j=0}^{3} \frac{\Gamma_j}{i g} \int_{q,p,\omega}  \mathcal{B}_\omega
\Bigl [1 + \frac{16 \Gamma_j i \omega}{g} \mathcal{D}_{\bm{q}+\bm{p}}^{(j) R}(\omega)  \Bigr ]   \mathcal{D}^{R2}_q(\Omega) \mathcal{D}^R_p(\omega+\Omega)
\notag \\
&  \hspace{0.5cm}
+ \left ( \frac{32}{g}\right )^2 \sum_{j=0}^{3} \frac{2 \Gamma_j}{i g} \int_{q,p,\omega} \Bigl [ 2 \mathcal{B}_\omega - \mathcal{F}_{\omega-E} - \mathcal{F}_{\omega+E^\prime}\Bigr ] \mathcal{D}^R_p(\Omega-\omega)
\mathcal{D}^{R2}_q(\Omega)\Bigl [1 + \frac{16 \Gamma_j i \omega}{g} \mathcal{D}_{\bm{q}+\bm{p}}^{(j) R}(\omega)  \Bigr ]
 \notag \\
&  \hspace{0.5cm}
+  \left ( \frac{64}{g}\right )^2  \frac{Z_\omega}{i g} \int_{q,p,\omega}
  \mathcal{L}_{\bm{q}+\bm{p}}^{R}(\omega) \mathcal{C}^{R2}_q(\Omega)
  \Bigl \{ \mathcal{F}_{\omega-E} \mathcal{C}^R_p(\omega-\mathcal{E})
  +\mathcal{F}_{\omega+E^\prime}  \mathcal{C}^R_p(\omega+\mathcal{E}) \bigr ]
  \Bigr \}
.
\label{eq2loopK2_P2+-1}
\end{align}
Here we again took into account that diffuson $\mathcal{D}^R_q(\omega)$ and cooperon $\mathcal{C}^R_q(\omega)$ propagators are the same. The most part of the two-loop contribution to $[P_2^{\alpha_1\alpha_2}]^{RA}(E,E^\prime)$ can be recast
in the form of the diffuson and cooperon one-loop contribution renormalized by interaction and disorder
(see Appendix \ref{app:sec:2}). In the limit $E=E^\prime=T=0$, we arrive (see details for evaluation of the integrals in Ref.~[\onlinecite{Burmistrov2015a}])
at
\begin{gather}
[P_2^{\alpha_1\alpha_2}]^{RA (2)} \to - 32 \frac{t^2\,h^{2\epsilon}}{\epsilon^2} \Bigl [ 3+\epsilon \Bigr ]- 16 \frac{t^2\,h^{2\epsilon}}{\epsilon^2}\Bigl [
2 \sum_{j=0}^3 f(\gamma_j )
+ 3 \sum_{j=0}^3 \ln (1+\gamma_j) +4\gamma_c
 -
\epsilon \sum_{j=0}^3 \frac{2+\gamma_j}{\gamma_j}  \Bigl ( \ln(1+\gamma_j)
\notag \\
+ \liq(-\gamma_j) + \frac{1}{4}\ln^2(1+\gamma_j)\Bigr ) +2 \epsilon \gamma_c\Bigr ],
\label{eq2loopK2_P2+-2}
\end{gather}
where the function $f(x)$ is given by Eq.~(\ref{fx}) of the main text and the function $\liq(x)$ is the polylogarithm.
We note that the result \eqref{eq2loopK2_P2+-2} is of the first order in $\gamma_c$. This occurs since the
terms of the first order in the fluctuation propagator exist in Eq. \eqref{eq2loopK2_P2+-1}.
Combining Eqs. \eqref{eq2loopK2_P2++2} and \eqref{eq2loopK2_P2+-2}, we arrive at Eq.~(\ref{eqK2_2}) of the main text.

\section{One-loop renormalization of (mesoscopic) diffuson and cooperon propagators}
\label{app:sec:2}

In this Appendix, we present the one-loop results for the renormalization of the (mesoscopic) diffuson and cooperon propagators.
Such renormalization accounts for the significant part of the two-loop contribution to $[P_2^{\alpha_1\alpha_2}]^{RA}(E,E^\prime)$.
Taking into account Eq. \eqref{eq18P1} and Eq. \eqref{eq2loopK2_P2+-1}, we can rewrite the expression for $[P_2^{\alpha_1\alpha_2}]^{RA}(E,E^\prime)$
in the following way:
\begin{gather}
[P_2^{\alpha_1\alpha_2}]^{RA}(E,E^\prime) = -  \int_q \frac{256 Z(E,E^\prime)}{g q^2 - 16 i Z_\omega \Omega - \Sigma^R(q,E,E^\prime)}
- 2 \left (\frac{16}{g} \right )^2 \left (\int_q \mathcal{D}^R_q(\Omega) \right )^2
- 4 \left (\frac{16}{g} \right )^2 \int_{qp} \mathcal{D}^{R2}_q(\Omega)  .
\end{gather}
Here the renormalization factor $Z(E,E^\prime)$ is given as (cf. Eq. \eqref{rhores})
\begin{align}
Z(E,E^\prime) & = 1 + \frac{16}{i g^2} \sum_{j=0}^3 \Gamma_j \int_{p,\omega} \Bigl [ \mathcal{F}_{\omega-E}+  \mathcal{F}_{\omega+E^\prime}\Bigr ]
\mathcal{D}^{R}_p(\omega)\mathcal{D}^{(j)R}_p(\omega)
\notag \\
& + \frac{32Z_\omega}{i g^2} \int_{p,\omega} \Biggl \{ \mathcal{C}^{R2}_p(2E-\omega)
\Bigl [ \mathcal{L}^{K}_p(\omega)+\mathcal{F}_{E-\omega} \mathcal{L}^{R}_p(\omega)\Bigr ]
+\mathcal{C}^{R2}_p(\omega-2E^\prime) \Bigl [ \mathcal{L}^{K}_p(\omega)+\mathcal{F}_{\omega-E^\prime} \mathcal{L}^{A}_p(\omega)\Bigr ]
\Biggr \}.
\end{align}
The diffuson self-energy reads
\begin{gather}
\Sigma^R(q,E,E^\prime)   = 4 q^2 \int_p \mathcal{D}^{R}_p(\Omega)
- \frac{8}{ig} \sum_{j=0}^3 \Gamma_j \int_{p,\omega}
\Bigl [2\mathcal{B}_\omega- \mathcal{F}_{\omega-E}-  \mathcal{F}_{\omega+E^\prime}\Bigr ]   \frac{\mathcal{D}^{(j)R}_p(\omega)}{\mathcal{D}^{R}_p(\omega)}
\Bigl [ \mathcal{D}^{R}_{\bm{p}+\bm{q}}(\omega+\Omega) + \mathcal{D}^{A}_{\bm{p}-\bm{q}}(\omega-\Omega) \Bigr ]
\notag \\
 + \frac{8}{ig} \sum_{j=0}^3 \Gamma_j \int_{p,\omega} \Bigl [ \mathcal{F}_{\omega-E}+  \mathcal{F}_{\omega+E^\prime}\Bigr ]\mathcal{D}^{(j)R}_p(\omega) \Bigl [ 2 \bm{p} \bm{q} \mathcal{D}^{R}_{\bm{p}+\bm{q}}(\omega+\Omega) + \bigl [\mathcal{D}^{R}_q(\Omega)\bigr ]^{-1}
\bigl[ \mathcal{D}^{R}_{\bm{p}+\bm{q}}(\omega+\Omega) - \mathcal{D}^{R}_p(\omega)\bigr ]
\Bigr ] \notag \\
-\frac{16 Z_\omega}{i g}  \bigl [\mathcal{C}^{R}_q(\Omega)\bigr ]^{-1}
\int_{p,\omega} \Biggl \{ \mathcal{C}^{R2}_p(2E-\omega)
\Bigl [ \mathcal{L}^{K}_p(\omega)+\mathcal{F}_{E-\omega} \mathcal{L}^{R}_p(\omega)\Bigr ]
+\mathcal{C}^{R2}_p(\omega-2E^\prime) \Bigl [ \mathcal{L}^{K}_p(\omega)+\mathcal{F}_{\omega-E^\prime} \mathcal{L}^{A}_p(\omega)\Bigr ]
\Biggr \}
\notag \\
- \frac{16 Z_\omega}{i g} \int_{p,\omega} \mathcal{L}^{R}_{\bm{p}+\bm{q}}(\omega)
\Bigl [ \mathcal{F}_{\omega-E} \mathcal{C}^{R}_p(\omega-\mathcal{E})+\mathcal{F}_{\omega+E^\prime} \mathcal{C}^{R}_p(\omega+\mathcal{E}) \Bigr ]
\notag \\
-\frac{16 Z_\omega}{i g} \int_{p,\omega}
\Biggl \{ \mathcal{C}^{R}_p(2E-\omega)
\Bigl [ \mathcal{L}^{K}_p(\omega)+\mathcal{F}_{E-\omega} \mathcal{L}^{R}_p(\omega)\Bigr ]
+\mathcal{C}^{R}_p(\omega-2E^\prime) \Bigl [ \mathcal{L}^{K}_p(\omega)+\mathcal{F}_{\omega-E^\prime} \mathcal{L}^{A}_p(\omega)\Bigr ]
\Biggr \}
.
\end{gather}
Expanding the self-energy $\Sigma^R(q,E,E^\prime)$ to the lowest order in $\omega$ and $q^2$, we find
\begin{equation}
\frac{1}{g} \mathcal{D}^R_q(\Omega) \quad  \to \quad \frac{Z(E,E)}{g(E) q^2 - i 16 Z_\omega(E) \Omega+16 Z_\omega(E) \tau_\phi^{-1}(E)} ,
\end{equation}
where
\begin{align}
g(E) & = g - 4 \int_p \mathcal{D}_p ^R(0)+ \frac{16}{g} \sum_{j=0}^3  \Gamma_j \int_{p,\omega}p^2
 \Bigl [\mathcal{F}_{\omega-E}+  \mathcal{F}_{\omega+E}\Bigr ] \im \Bigl [ \mathcal{D}^{(j) R}_p(\omega) \mathcal{D}^{R2}_{p}(\omega)\Bigr ] \notag \\
 & +\frac{64 Z_\omega}{g} \int_{p,\omega}p^2 \mathcal{F}_{\omega-E}
 \im \Bigl [ \mathcal{L}^{R}_{p}(\omega)\mathcal{C}^{R3}_{p}(\omega-2E)\Bigr] \notag \\
 & -\frac{16}{g} \sum_{j=0}^3  \Gamma_j \int_{p,\omega}
 \Bigl [2 \mathcal{B}_\omega - \mathcal{F}_{\omega-E}-  \mathcal{F}_{\omega+E}\Bigr ] \im \Bigl [ \mathcal{D}^{(j) R}_p(\omega) [\mathcal{D}^{R}_{p}(\omega)]^{-1}\Bigr ] \re \Bigr [ [1-2p^2\mathcal{D}^{R}_{p}(\omega)] \mathcal{D}^{R2}_{p}(\omega)\Bigr ] \notag \\
 & + \frac{64 Z_\omega}{g} \int_{p,\omega} \Bigl [\mathcal{B}_\omega+ \mathcal{F}_{E-\omega}\Bigr ]
 \im \mathcal{L}^{R}_{p}(\omega) \re \mathcal{C}^{R2}_{p}(\omega-2E)
 , \label{eqgE} \\
Z_\omega(E) & = Z_\omega + \frac{1}{2 g} \sum_{j=0}^3 \Gamma_j  \int_{p,\omega} \partial_\omega\Bigl [\mathcal{F}_{\omega-E}+  \mathcal{F}_{\omega+E}\Bigr ] \re \Bigl [ \mathcal{D}^{(j)R}_p(\omega) [\mathcal{D}^R_{p}(\omega)]^{-1} \Bigr ] \re \mathcal{D}^R_{p}(\omega)
\notag \\
& +\frac{128 Z_\omega^2}{g^2}  \int_{p,\omega} \Bigl [\mathcal{B}_\omega+ \mathcal{F}_{E-\omega}\Bigr ]
 \im \mathcal{L}^{R}_{p}(\omega) \re \mathcal{C}^{R2}_{p}(\omega-2E)
 \notag \\
 &+ \frac{4Z_\omega}{g} \int_{p,\omega}\partial_\omega \mathcal{F}_{\omega-E}
 \re \Bigl [\mathcal{L}^{R}_{p}(\omega) \mathcal{C}^{R}_{p}(\omega-2E)\Bigr ]
 \notag \\
&+ \frac{2Z_\omega}{g} \int_{p,\omega}\partial_\omega \mathcal{F}_{\omega-E}
 \re \Bigl [\mathcal{L}^{R}_{p}(\omega) \mathcal{C}^{A}_{p}(\omega-2E)\Bigr ]
 \notag \\
 &+ \frac{4Z_\omega}{g} \int_{p,\omega}\mathcal{F}_{\omega-E}
 \re \Bigl [\partial_\omega \mathcal{L}^{R}_{p}(\omega) \mathcal{C}^{R}_{p}(\omega-2E)
 -\mathcal{L}^{R}_{p}(\omega) \partial_\omega \mathcal{C}^{R}_{p}(\omega-2E)\Bigr ]
  , \label{eqzE} \\
\tau_\phi^{-1}(E) & =\frac{4}{g} \sum_{j=0}^3 \Gamma_j  \int_{p,\omega}\Bigl [2\mathcal{B}_\omega- \mathcal{F}_{\omega-E}-  \mathcal{F}_{\omega+E}\Bigr ]  \im \Bigl [ \mathcal{D}^{(j)R}_p(\omega) [\mathcal{D}^R_{p}(\omega)]^{-1} \Bigr ] \re \mathcal{D}^R_{p}(\omega) \notag \\
& - \frac{4}{g} \int_{p,\omega}\Bigl [ \mathcal{B}_\omega+ \mathcal{F}_{E-\omega} \Bigr ] \im \mathcal{L}^R_{p}(\omega)
\re \mathcal{C}^R_{p}(\omega-2E) . \label{eqTE}
\end{align}
As one can see, for $|E|\gg T$, the energy $|E|$ indeed serves as the cut-off for the infrared logarithmic divergences in Eqs. \eqref{eqgE} and \eqref{eqzE} (in the case of $L\to \infty$). At the same time, the non-zero value of energy $E$ induces finite dephasing time.

We note that Eqs. \eqref{eqgE} and \eqref{eqzE} reproduce the one-loop renormalization of $g$ and $Z_\omega$, respectively, as it was found in
Ref.~[\onlinecite{Burmistrov2015b}] with the help of the background field method. The Cooper-channel contribution to the dephasing rate (the second line in Eq. \eqref{eqTE}) coincides with the result found in Refs. [\onlinecite{BrenigWoelfle1985},\onlinecite{Reizer1992}].


\section{The average local density of states}
\label{Sec:App:LDOS}

In this Appendix, we present details of a
perturbative analysis of the average local density of states.
We start from rewriting Eq. \eqref{rhores} in the following form
\begin{equation}
\langle \rho(E) \rangle = \rho_0 + \delta\rho_{\rm ph}(E)+\delta\rho^{(1)}_{\rm pp}(E) +\delta\rho^{(2)}_{\rm pp}(E),
\end{equation}
where
\begin{gather}
\frac{\delta\rho_{\rm ph}(E)}{\rho_0}
=   \im  \sum_{j=0}^3 \frac{16 Z_\omega \gamma_j}{g^2} \int_{q,\omega}  \mathcal{F}_{\omega-E}
\mathcal{D}^R_q(\omega) \mathcal{D}^{(j) R}_q(\omega) ,
\label{rhores-app1}
\end{gather}
\begin{gather}
\frac{\delta\rho^{(1)}_{\rm pp}(E)}{\rho_0}   =  \im \int_{q,\omega} \frac{32 Z_\omega}{g^2}  \mathcal{F}_{E-\omega}\mathcal{C}^{A2}_q(\omega-2E)  \mathcal{L}^A_q(\omega) ,
\label{rhores-app2}
\end{gather}
and
\begin{gather}
\frac{\delta\rho^{(2)}_{\rm pp}(E)}{\rho_0}    =  \re \int_{q,\omega}  \frac{64 Z_\omega}{g^2} \mathcal{C}^{R2}_q(2E-\omega) \Bigl [
 \mathcal{B}_\omega +\mathcal{F}_{E-\omega} \Bigr ]
  \im \mathcal{L}^R_q(\omega) .
\label{rhores-app3}
\end{gather}

\subsection{Fluctuation corrections above $T_c^*$ at criticality}

For temperatures $T$ close to $T_c^*$, $T-T_c^*\ll T_c^*$, the most important contribution comes from the term $\delta\rho^{(2)}_{\rm pp}(E)$. In this case, the integrals over frequency and momentum are dominated by region $Dq^2, |\omega| \ll T_c^*$. In this case, the fluctuation propagator \eqref{eq:prop:fp} can be written in the following form:
\begin{equation}
\mathcal{L}_q^R(\omega) = - \frac{8 T_c^*}{\pi} \frac{1}{\tau_{GL}^{-1} + D q^2 - i \omega} ,
\label{eq:app:FLP}
\end{equation}
where $\tau^{-1}_{GL} = 8 T_c^* |\gamma_c^{-1}(L_T)|/\pi \ll T_c^*$. For energies $|E|\ll T_c^*<T$, taking into account that the RG flow is stopped at the length scale $L_T$, we find
\begin{equation}
\frac{\delta\rho^{(2)}_{\rm pp}(E)}{\rho_0}  = - t(L_T) (T_c^* \tau_{GL})^\frac{6-d}{2} \mathcal{H}_d(2 |E| \tau_{GL}) .
\label{eq:app:1:1}
\end{equation}
Here the function $\mathcal{H}_d(z)$ is given by
\begin{equation}
\mathcal{H}_d(z) =
\frac{8}{\pi} \int\limits_0^\infty dy  \int \limits_{-\infty}^{\infty} \frac{dx\, y^\frac{d-2}{2}}{(1+y)^2+x^2} \re \frac{1}{(y+ix-2iz)^2}
\end{equation}
We note that $t(L_T)$  can be approximated by $t_\ast$ in the critical region, $\delta_\xi\ll T_c^*$.
Hence from Eq. \eqref{eq:app:1:1} for $|E|\ll \tau_{GL}^{-1}$ we find Eq.
\eqref{eq:1:e11:d} with $c_d = \mathcal{H}_d(0)$. We note that $c_2 = 8(1-\ln 2)$ (see Eq. \eqref{eq:1:e11a})and $c_3= 2\pi (3\sqrt{2}-4)$ in agreement with results of Refs. [\onlinecite{Abrahams1970},\onlinecite{Maki1970},\onlinecite{Castro1990}]. At $|z|\gg 1$ the function $\mathcal{H}_d$ has the following asymptotic behavior: $\mathcal{H}_d(z)\approx - \tilde{c}_d z^\frac{d-6}{2}$, where
\begin{equation}
\tilde{c}_d = -
\frac{2^\frac{d}{2}}{\pi} \int\limits_0^\infty dy  \int \limits_{-\infty}^{\infty} \frac{dx\, y^\frac{d-2}{2}}{(1+x)^2+y^2} \re \frac{y^2-x^2}{(y^2+x^2)^2} .
\label{eq:app:tildecd}
\end{equation}
This asymptote leads to the result given by Eq.
\eqref{eq:1:e11:d} for energies $\tau_{GL}^{-1}\ll |E| \ll T_c^*$. We note that $\tilde{c}_3 = \pi/\sqrt{2}$, in agreement with Ref. [\onlinecite{Castro1990}]. We also note that the integral in Eq. \eqref{eq:app:tildecd} is logarithmically divergent in $d=2$. In this case, one finds $\tilde{c}_2 = 2\ln z$ (see Eq.  \eqref{eq:1:e11b}).

Finally, we consider energies $|E| \gg T_c^*$. The dominant contribution comes from the region $Dq^2, |\omega| \ll T_c^*$ in which the fluctuation propagator can be written in the form \eqref{eq:app:FLP}. Then, we obtain
\begin{equation}
\frac{\delta\rho^{(2)}_{\rm pp}(E)}{\rho_0}  = \frac{4 t(L_E)}{\pi} \left (\frac{T_c^*}{E}\right )^2
\int\limits_{0}^{\sim 1}d y \int\limits_{0}^{\sim 1} \frac{dx\, y^\frac{d-2}{2} }{(y+ \frac{1}{T_c^*\tau_{GL}})^2+x^2}
\label{eq:app:1:2}
\end{equation}
Hence for $d>2$ the fluctuation correction to the average local density of states at energies $|E| \gg T_c^*$ is proportional to $t(L_E) ({T_c^*}/{E})^2$.
In $d=2$ the integral on the right-hand side of Eq. \eqref{eq:app:1:2} diverges logarithmically and one obtains Eq. \eqref{eq:1:e11a}.

\subsection{The average local density of states in the insulating phase at $T=0$}

In this section, following the approach proposed in Ref.~[\onlinecite{Burmistrov2014}], we evaluate the average LDOS in the insulating phase.
In the insulating phase (the region $\delta_\xi\gg T_c^*$ and $t_0>t_\ast$), the conductivity can be written in the scaling form:
\begin{equation}
g(q,\omega) = \xi^{2-d} \mathcal{R}_D\left ({\omega}/{\delta_\xi}, q\xi\right ) .
\label{eq:app:g:sf}
\end{equation}
The asymptotic behavior of the scaling function $\mathcal{R}_D$ is summarized in the Table \ref{Tab_IntDiffD}. Equation~\eqref{eq:app:g:sf} implies the following scaling form for the diffusion coefficient $D=g/16Z_\Omega$:
\begin{equation}
D(q,\omega) = \xi^2 \left ({\xi}/{l}\right )^{-d} E_0 \mathcal{R}_D\left ({\omega}/{\delta_\xi}, q\xi\right ) .
\end{equation}
Here we have introduced the ultraviolet energy scale $E_0=1/(16 Z_\omega l^d)$. In what follows, we
assume that $t(\xi) \sim t_\ast \sim 1$ and consider the case of BCS line $\gamma_t = -\gamma_s = \gamma_c=\gamma$.
We remind the reader that in the presence of attraction the insulating phase occurs at $\delta_\xi \gg T_c^*$ which
implies $|\gamma(\xi)|\ll 1$.

To the lowest order in $\gamma$, the contributions of the RG type from the particle-hole and particle-particle channel can be summed and written as
\begin{align}
\frac{\delta\rho_{\rm ph}(E)+\delta\rho_{\rm pp}^{(1)}(E)}{\rho_0} & =
16 \Omega_d \im\int \limits_{\varepsilon}^\infty d\Omega  \int \limits_0^\infty d Q \,Q^{d-1}
\frac{ \gamma(\Omega, Q)}{(R_D(\Omega,Q) Q^2 - i \Omega)^2} .
\label{eq:app:g:sf2}
\end{align}
Here $\gamma(\Omega,Q) = \gamma(\xi) (\max\{Q^{-1}, L_\Omega\})^{|\Delta_2|}$ and
we have defined $\varepsilon = E/\delta_\xi$. For $|\Omega| \gg 1$ the
integral over $Q$ in Eq. \eqref{eq:app:g:sf2} is dominated by $Q \sim |\Omega|^{1/d}$.
Then we obtain
\begin{equation}
\im\int \limits_0^\infty d Q \,Q^{d-1}  \frac{ \gamma(\Omega, Q)}{(R_D(\Omega,Q) Q^2 - i \Omega)^2} \sim \frac{\gamma(L_\Omega)}{\Omega} .
\end{equation}
Hence we reproduce RG result \eqref{eq:LDOS:CR1} for $\varepsilon\gg 1$ from Eq. \eqref{eq:app:g:sf2}.

\begin{table*}[t]
\caption{The asymptotic behavior of the scaling function $\mathcal{R}_D(\Omega,Q)$. The scaling of $R_D$ for $|\Omega| \ll 1$ is written in accordance with the Mott's formula. Here $c$ is a positive constant of the order unity.}
\label{Tab_IntDiffD}
\begin{tabular}{l|lc|lc}
& & $0\leqslant |\Omega|\ll 1$ & & $1\ll |\Omega|$ \\
\hline $\max\{|\Omega|^{1/d}, 1\} \ll Q$ & (I): &
$Q^{d-2+\Delta_2}(-i\Omega + c \Omega^2 \ln^{d+1}(1/|\Omega|))$
 & (III): &  $Q^{d-2+\Delta_2} |\Omega|^{-\Delta_2/d}$ \\
$0\leqslant Q \ll \max\{|\Omega|^{1/d}, 1\}$& (II): &  $-i\Omega + c \Omega^2 \ln^{d+1}(1/|\Omega|)$& (IV): & $|\Omega|^{(d-2)/d}$\\
\end{tabular}
\end{table*}

For $|\Omega| \ll 1$ the integral over $Q$ in Eq. \eqref{eq:app:g:sf2} is dominated by $Q\sim 1$. Then
we find
\begin{equation}
\im\int \limits_0^\infty d Q \,Q^{d-1}  \frac{ \gamma(\Omega, Q)}{(R_D(\Omega,Q) Q^2 - i \Omega)^2} \sim \frac{\gamma(\xi)}{\Omega} \ln^{d+1} \frac{1}{|\Omega|} .
\end{equation}
Then integrating over $\Omega$ in Eq. \eqref{eq:app:g:sf2}, we obtain
\begin{equation}
\frac{\delta\rho_{\rm ph}(E)+\delta\rho_{\rm pp}^{(1)}(E)}{\rho_0}= a_1 \gamma(\xi)  \ln^{d+2} \frac{1}{|\varepsilon|} ,
\end{equation}
where $a_1$ is some positive constant.

The other contribution from the particle-particle channel is of non-RG type.
For $|\gamma(\xi)|\ll 1$ this term becomes of the second order in $\gamma(\xi)$:
\begin{gather}
\frac{\delta\rho_{\rm pp}^{(2)}(E)}{\rho_0}  =
16 \Omega_d \re\int \limits_0^{\varepsilon} d\Omega  \int \limits_0^\infty \frac{d Q \,Q^{d-1}}{(R_D(\Omega,Q) Q^2 - i \Omega)^2}
 \im \frac{1}{\gamma^{-1}(\xi) - \ln(R_D(\Omega,Q) Q^2 - i \Omega)} \notag \\
 \approx - 8 \pi \Omega_d\gamma^2(\xi) \re\int \limits_0^{\varepsilon} d\Omega  \int \limits_0^\infty \frac{d Q \,Q^{d-1}}{(R_D(\Omega,Q) Q^2 - i \Omega)^2} .
 \label{eq:app:g:sf3}
\end{gather}
For  $|\Omega| \gg 1$ the integral over $Q$ in Eq. \eqref{eq:app:g:sf3} is dominated by $Q\sim |\Omega|^{1/d}$:
\begin{equation}
\int \limits_0^\infty \frac{d Q \,Q^{d-1}}{(R_D(\Omega,Q) Q^2 - i \Omega)^2} \sim \frac{1}{\Omega}
\end{equation}
Hence, for $\varepsilon \gg 1$ we find
\begin{equation}
\frac{\delta\rho_{\rm pp}^{(2)}(E)}{\rho_0}  \sim \gamma^2(\xi) \ln \varepsilon .
\end{equation}
This contribution is smaller than the RG result \eqref{eq:LDOS:CR1}.

For $|\Omega|\ll 1$  the integral over $Q$ in Eq. \eqref{eq:app:g:sf3} is dominated by $Q\sim 1$,
provided the inequality  $d+2\Delta_2>0$ holds. Then we obtain
\begin{equation}
\frac{\delta\rho_{\rm pp}^{(2)}(E)}{\rho_0} =  a_2 \gamma^2(\xi) \re \int \limits_0^{\varepsilon} \frac{d\Omega}{(\Omega+i0)^2} =  - a_2 \gamma^2(\xi)  \frac{1}{\varepsilon} ,
\end{equation}
where $a_2$ stands for a positive constant.

\end{widetext}

\bibliography{paper}

\begin{thebibliography}{58}
\bibitem [{ftn()}]{ftn1}%
  \BibitemOpen
  \href@noop {} {}\bibinfo {note} {{To make the plots we change $1 - 8(1-\ln 2)
  t_c (T_c \tau_{GL})^2 $ in Eq. \eqref{eq:1:e11a} to $1/[1 + 8(1-\ln 2) t_c
  (T_c \tau_{GL})^2 ]$.}}\BibitemShut {Stop}%
\bibitem [{not()}]{note-samuely}%
  \BibitemOpen
  \href@noop {} {}\bibinfo {note} {{The authors of Ref.~[\onlinecite{Szabo16}]
  observed neither a depletion of LDOS around zero energy above $T_c$ nor
  sizeable spatial fluctuations of LDOS in MoC films. This can be presumably
  attributed to the fact that all the measurements in
  Ref.~[\onlinecite{Szabo16}] were carried out quite far from SIT (on its
  superconducting side). Indeed, the largest value of the resistance maximum
  ($t_c$) in this work was 1.4 k$\Omega$, i.e., twenty times lower than the
  quantum value $h/e^2$. We expect that if measurement would be conducted on
  more resistive samples (closer to the SIT as well as on the insulating side
  of SIT), all the effects analyzed in our paper (and observed in TiN, InO, and
  NbN films) would emerge.}}\BibitemShut {Stop}%
\end{thebibliography}

\end{document}